\documentclass[%
 aapm,
 mph,%
 amsmath,amssymb,
 twocolumn
]{revtex4}

\usepackage{graphicx}
\usepackage{dcolumn}
\usepackage{bm}
\usepackage{epsf}

\usepackage[mathlines]{lineno}
\modulolinenumbers[5]
\renewcommand{\vec}[1]{\mbox{\boldmath$ #1 $}}

\begin{document}

\preprint{AAPM/123-QED}

\title{Self-organisation of spermatozoa in bulk suspensions\\via unsteady elastohydrodynamic interactions}

\author{Nanami Taketoshi\footnote{Address: Aramaki Aoba 6-6-01, Sendai, Miyagi, Japan}}
\email{nanami.taketoshi.q1@dc.tohoku.ac.jp}

\author{Toshihiro Omori}
\email{omori@tohoku.ac.jp}

\author{Takuji Ishikawa}
\affiliation{Department of Biomedical Engineering, Tohoku University}
\affiliation{Department of Finemechanics, Tohoku University}
\date{\today}

\begin{abstract}
We developed a mechanical model of spermatozoal swimming in bulk suspensions. We traced the spatiotemporal elastohydrodynamic interactions and found that spermatozoa engaged in self-organisation: flagellar undulatory motion generated a stable polar order. Conventional microswimmer theories use time-averaged stress fields to predict that the cellular structure is unstable. However, our present results reveal a novel self-organisation mechanism attributable to unsteady hydrodynamic interactions. We thus contribute to the theory of active soft matter dynamics.
\end{abstract}

\maketitle
Spermatozoal motility is essential for reproduction and is increasingly important given the rising human infertility rates in developed countries \cite{Fauci,Gaffney}.
Spermatozoal swimming is driven by the undulatory motions of flagella that generate a time-variant fluid flow around each cell \cite{Chakrabarti,Fauci2,IshimotoPRL,Smith}.
Spermatozoa have been conventionally categorised as pushers at low Reynolds numbers \cite{IshimotoPRL,Saintillan}; such theories generally employ time-averaged flows attributable to flagellar beating.
The active stress created by pushers renders flow unstable \cite{Miles,Ishikawa}, and often induces turbulence-like swarming \cite{Creppy,IshikawaPRL} during collective swimming of small groups \cite{Ishimoto3,Schoeller,Tung}. However, pushers form coherent structures in limited situations, such as within closed spaces \cite{Lushi}.
The importance of the spatiotemporal interactions between flow fields and cell motility has recently been emphasised \cite{OmoriJFM}. For example, local interactions among swimmers that assume helical trajectories generate polar order \cite{Samatas}. 
Spermatozoal flagellar waveforms can vary by the physical environment \cite{Smith,OmoriMaxwell,Rikmenspoel}. Spatiotemporal variations in flagellar motion may result in hydrodynamic flagellar synchronisation \cite{Camalet,ChakrabartiPRL,Goldstein,Oriola,Riedel-Kruse,Sartori,Ramin} or rheotaxis \cite{Miki,Kantsler,Ishimoto2,OmoriPRE}.

Here, we show that spermatozoa form a polar order in a bulk suspension.
The spatiotemporal elastohydrodynamic interactions in a suspension are analysed by developing a model that explicitly represents the elastic deformation of spermatozoal flagella, the driving force of dynein, and interactions with the intercellular fluid.
Stable polar order is achieved irrespective of the number density, but the time required for establishment of the order decreases with density.
Individual spermatozoa spontaneously assume a stable structure via repeated three-dimensional elastohydrodynamic interactions in a system with six degrees of freedom.
To elucidate the mechanism by which the coherent structure forms, we compare our model to a coarse-grained model in which individual spermatozoa are assumed to be steady pushers, thus swimmers, the active stress generated by steady pusher does not vary over time. 
We show that polar order is not achieved by steady pushers but rather is formed via unsteady fluid-structure interactions.
Our results suggest that the global structure formed via time-averaged interactions is quite different from that formed by ever-changing interactions, indicating the importance of unsteady interactions in collective active soft matter dynamics.

{\it Modelling of spermatozoa and the problem setting--}
Consider the elastohydrodynamic interactions of spermatozoa swimming within their suspensions.
The governing equation and the numerical methods used can be found in the supplemental material and our former studies \cite{Ito,OmoriPNAS,Taketoshi}, and are thus only briefly described here.

Assume that motile spermatozoa swim freely in a Newtonian liquid without walls.
Given their small size, the effect of fluid inertia is negligible and fluid motion can be assumed to be governed by the Stokes equation.
The flow field is then expressed by a boundary integral equation of slender body theory \cite{Tornberg} and is solved using a boundary element method \cite{Ito,OmoriPNAS}.

The flagellum of eukaryotic spermatozoa features a 9+2 doublet microtubule and is elastically deformable.
The mean radius $a$ of a spermatozoal flagellum is very small compared to its length $L$ (e.g., $a/L \simeq 0.002$ for human spermatozoa \cite{Gaffney}) and the transverse shear stress of flagellar deformation is thus assumed to be negligible.
The Euler-Bernoulli assumption is adopted when considering elastic flagellar deformation; the flagellum is modelled as a homogeneous, isotropic elastic body.
The solid mechanics of flagellar deformation are solved using a finite element method and we simulate the fluid-structure interactions employing a boundary element-finite element coupling method \cite{Taketoshi}.

Here, we model the driving forces of the molecular motor, dynein, that powers the flagellar beats.
Assume that a balance is in play among the active force $\vec{q}_a$, the fluid viscous force $\vec{q}_f$, and the elastic force $\vec{q}_e$: $\vec{q}_a + \vec{q}_f + \vec{q}_e = \vec{0}$. 
A reference $\vec{q}_a$ waveform\cite{Taketoshi} is input for all spermatozoa (but the phase varies individually) and $\vec{q}_e$ is calculated from the flagellar shape using a finite element method.
By determining $\vec{q}_f$ by reference to the force balance, the flagellar velocity is given by the boundary integral equation.
Although the reference $\vec{q}_a$ is common to all spermatozoa, the flagellar waveform may vary via elastohydrodynamic interaction among individual flagella.

Spermatozoa are immersed in a cubic domain of length $L_c = 4.0L$, where $L$ is the length of a flagellum.
To apply triply periodic boundary conditions, we employ the Ewald summation of the Beenakker method \cite{Beenakker}.
The number of spermatozoa in the domain varies from $N = 72$, to 144, to 216.
A typical human spermatozoal flagellum is about 60 $\mu$m \cite{Gaffney} in length, and the cell numbers thus translate into number densities of 5, 10 and 15 million cells/mL. For comparison, the spermatozoal density of normal human semen is over 15 million/mL.

\begin{figure}[h]
\resizebox{0.8\textwidth}{!}{
\includegraphics[]{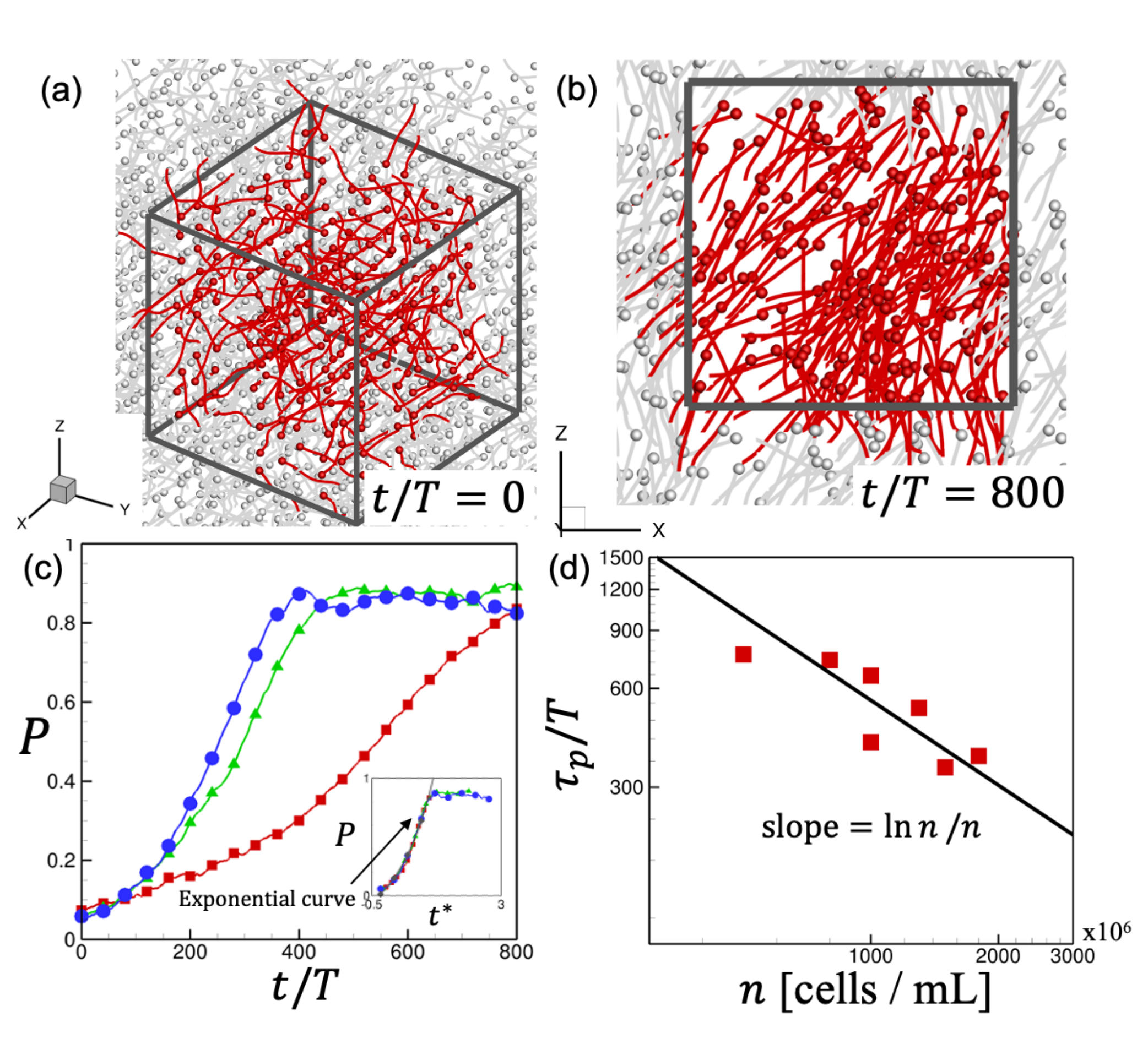}}
\caption{Self-organisation of swimming spermatozoa ($N=216$). (a) Spermatozoa are initially randomly aligned and (b) after a long time, become re-oriented in polar order. (c) The ensemble average swimming direction $P$. Red, green, and blue lines indicate $N = 72$, 144, and 216, respectively. In the inset, the data are normalised by the relaxation time $\tau_p$; all curves coincide with a single exponential curve. (d) Relaxation time as a function of number density. The slope $\ln n/n$ is also plotted. $T$ is the flagellar beating period.}
\label{fig1}
\end{figure}

{\it Self-organisation via elastohydrodynamic interactions--}
First, we investigate spermatozoal dynamics under high-density conditions ($N=216$).
To avoid among-spermatozoa contact, the spermatozoa are initially arranged in a grid with small fluctuations, but the swimming directions and flagellar beat phases are random (cf. Fig.\ref{fig1}a, henceforth termed ‘of random order’).
After the simulation commences, spermatozoa swim in the initially defined random directions, and coherent cell structures are therefore not observed (the average swimming direction $P =|\left<\vec{e}_i\right>|$ is about 0.1).
The bra-ket notation indicates the ensemble average of a physical quantity: $\frac{1}{N}\sum_i^N \cdot$.
The spermatozoa repeatedly engage in mutual interactions, gradually change their swimming directions, and after a long period (400 cycles), a polar order is observed; all cells swim in almost the same direction ($P \simeq 0.8$). This is maintained until the calculation ends at 800 cycles (Figs.\ref{fig1}b, c and Movie S1).

Does such self-organisation develop only above a certain cell number density threshold, and does a phase change accompany a density variation?
When the number density changes, self-organisation is observed even below the physiological conditions associated with male infertility (e.g., $N=72$, equivalent to $5 \times 10^{6}$ sperm/mL, in Fig.\ref{fig1}c).
However, the time required for re-orientation tends to decrease with increasing number density, and the relaxation time is density dependent.
Define the relaxation time $\tau_p$ as the time required to form the polar order, thus the time to when $P$ attains 0.8, and, after normalising the time variation of $P$ by $\tau_p$, the same exponential curve is obtained for all cases (see inset Fig.\ref{fig1}c).
Therefore, spermatozoal self-organisation does not evidence a phase change as the number density increases.
The time scale over which the polar order emerges is next estimated (the details are in the supplement).
The average $dP/dt$ is proportional to the number density $n$ (Fig.S5 in the supplement), suggesting that $\tau_p$ is basically proportional to $n^{-1}$.
In addition, the expected value of $P(t=0)$ given a random orientation of $N$ spermatozoa is $1/\sqrt{N}$\cite{Evans}, $\tau_p$ is eventually scaled to $\ln n/n$, in good agreement with the simulation results (cf. Fig.\ref{fig1}d).

\begin{figure}[h]
\resizebox{0.8\textwidth}{!}{
\includegraphics[]{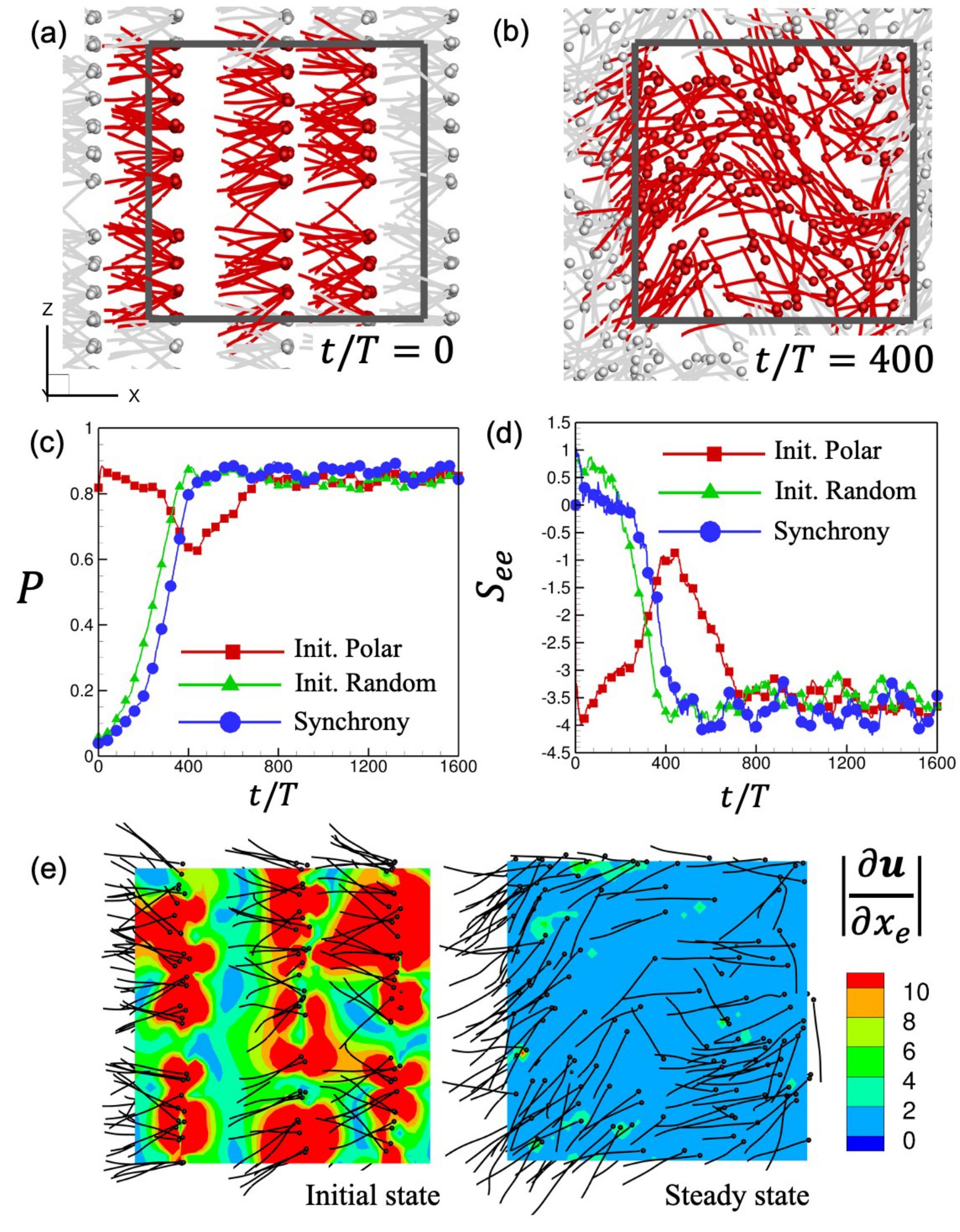}}
\caption{Spermatozoa initially aligned in polar order. (a, b) Changes over time in the spermatozoan configurations, thus at $t/T=0$,and 400. (c) The average swimming directions $P$ under three different conditions; thus initially random, initially polar, and with autonomous synchronisation.
(d) Stresslet changes over time during 1,600 beats. (e) Velocity gradient in the mean swimming direction initially (left) and in the steady state (right).}
\label{fig2}
\end{figure}

{\it Stability of the polar order structure--}
Although we found stable polar ordering, it has been suggested that coherent pusher structures are not maintained for long periods because of flow instability \cite{Alert,Ramaswamy,Ishikawa2009,IshikawaAPL}.
What factors stabilise the polar order?
To answer this question, the initial configuration is varied to be near-polar in order.
When spermatozoa are initially aligned in the same direction and the beating rhythm is almost the same (Fig.\ref{fig2}a), waving instability \cite{Ishikawa} develops (cf. Fig.\ref{fig2}b) and the mean swimming direction $P$ decreases from 0.8 to around 0.6 (at $t/T = 400$ in Fig.\ref{fig2}c).
Although the coherent cell structure is temporarily disrupted, the swimming direction is again re-oriented to the polar state and is maintained for a very long time, as revealed by the random order results (cf. Fig.\ref{fig2}c).
Although these results indicate that spermatozoal suspensions are stable in a polar order, they also suggest that different energetic states may exist for the same $P$-value. 

Are these different swimming states attributable to changes in the stresslet?
The ensemble average of the stresslet in the swimming direction $S_{ee}$ is shown in Fig.\ref{fig2}d.
When the spermatozoa attain a stable state, the value of $S_{ee}$ converges to about $-4$, and this is independent of the initial configuration.
Notably, the convergence value is equivalent to the initial value of the polar order instability, indicating that cell structure stability cannot be based on the stresslet alone.
We next investigated flow disturbance in the suspension, and the result is shown in Fig.\ref{fig2}e.
To assess the disturbance, we define the magnitude of the velocity gradient in the mean swimming direction as $|\partial \vec{u}/\partial x_e|$.
Initially, spermatozoa are aligned in the same swimming direction, and the beating surfaces are near-identical. Thus, fluid motions around cells are greatly disturbed and a high $|\partial \vec{u}/\partial x_e|$ region can be seen.
Spermatozoa swim with six degrees of freedom in three-dimensional space; spermatozoa can spontaneously adopt an arrangement that is stable in terms of both the beating surface and the inter-spermatozoal distance, reducing flow disturbance in the steady state even at the same $P$ and $S_{ee}$.

{\it Effect of self-synchronisation--}
We next investigate the effect of flagellar synchronisation on collective swimming.
We approach self-synchronisation via fluid motion. We thus add a curvature-dependent phase change model to the simulation, guided by \cite{Camalet,Chakrabarti,ChakrabartiPRL,Goldstein,Oriola,Riedel-Kruse,Sartori} (see supplement and \cite{Taketoshi} for the details).
The model is based on the geometric clutch hypothesis \cite{Lindemann}. The association-dissociation rate of the molecular motor dynein is regulated by the distance between the neighbouring microtubule, and the wave propagation of the active driving force is controlled by the waveform.
Autonomous flagellar synchronisation is achieved by changes in both the flagellar waveform and frequency attributable to elastohydrodynamic interactions. For example, flagellar synchronisation of two fixed spermatozoa is shown in Fig. 3a.

\begin{figure}[h]
\resizebox{0.8\textwidth}{!}{
\includegraphics[]{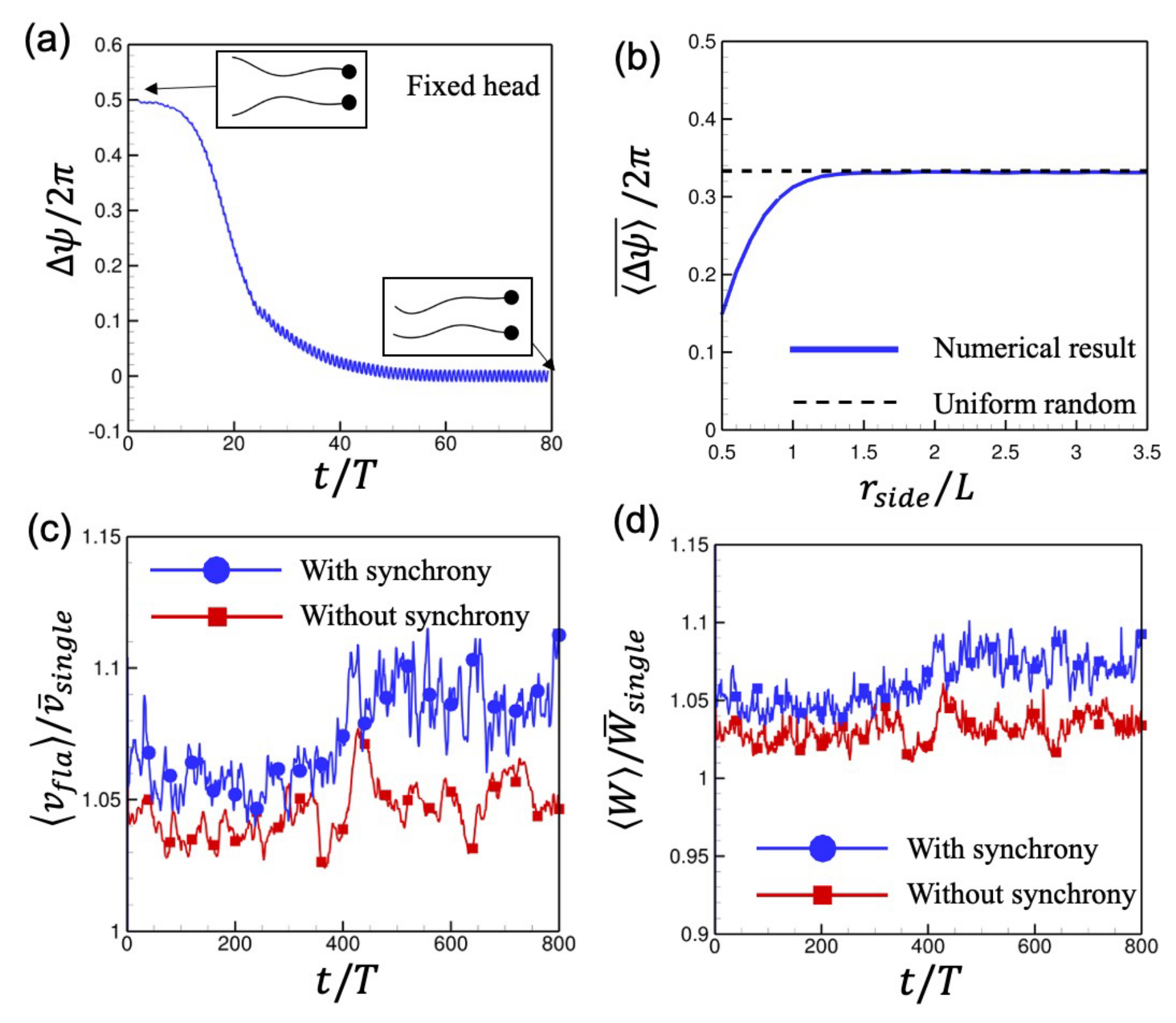}}
\caption{Self-synchronisation via elastohydrodynamic interactions.
(a) Changes over time in the phase difference between two fixed spermatozoa (initial phase difference $\pi$). Temporal snapshots of two flagella are also shown.
(b) Distance correlations between the phase differences in a spermatozoal suspension in the steady state (five-period average). If the differences were uniformly random, the expected value is one third. 
(c) Flagellar velocity and (d) mechanical work rate with or without synchronisation. 
These values are normalised to those of solitary swimming.} 
\label{fig3}
\end{figure}

To derive the phase difference in the suspension, the distance correlation difference is defined as $\left<\Delta \psi (r_{side},t)\right> = \frac{1}{N}\sum_{i}^{N}\left(\frac{1}{N_i^{side}}\sum_{j}^{N_i^{side}}|\psi_i - \psi_j|\right)$, where $r_{side}$ is the distance between a spermatozoon perpendicular to the swimming direction of the $i$th sperm and $N_i^{side}$ is the number of spermatozoa within $r_{side}$ from the $i$th spermatozoon.
Even in the free-swimming condition, the value is below the average that is expected (thus one third; see the supplement), suggesting that flagellar synchronisation is occurring where $r_{side}< L$ (cf. Fig.\ref{fig3}b).
The changes in the cell structure and the stresslet are small compared to the extent of flagellar synchronisation, which constitutes a stable polar order (see Figs.\ref{fig2}c, d).
The flagellar beating synchrony affects the flagellar dynamics more than does flagellar structure, increasing the beating speed and the work rate of spermatozoa when swimming (Figs.\ref{fig3}c, d).
These results are caused principally by frequency changes; there is little change in the mean beat amplitude on application of the synchronisation model (Fig.S12 in the supplement).

\begin{figure}[h]
\resizebox{0.6\textwidth}{!}{
\includegraphics[]{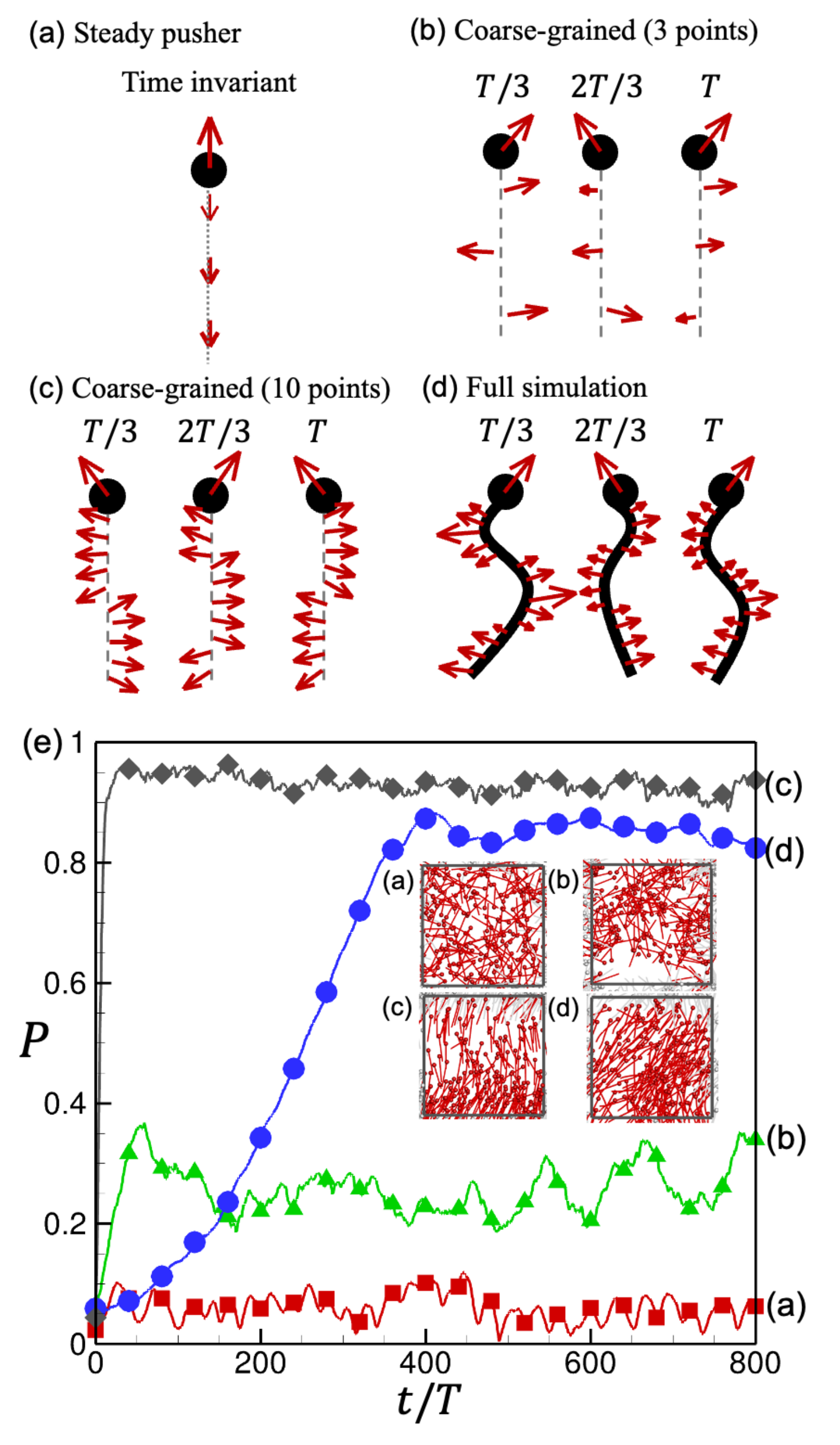}}
\caption{ Mechanisms by which polar orders emerge. A comparison of four different swimmers. (a) A steady pusher; the time-averaged driving force acts on the head and the counter-forces are distributed along the swim vector axis to satisfy the force-free condition. (b, c) A coarse-grained undulatory swimmer; the time-varying driving forces are applied to representative points (three or ten). The driving force dataset was created by reference to the full simulation. (d) The full simulation model, thus equivalent to Fig. 1. (e) Changes over time in the average orientations of all models. Snapshots taken at $t = 800T$ are also plotted. 
}
\label{fig4}
\end{figure}

{\it Mechanism of structure formation --}
Finally, we discuss how spermatozoa self-organise.
We built coarse-grained models of spermatozoal swimming to investigate the effects of unsteady flagellar beating.
The first model features a steady pusher (Fig.\ref{fig4}a) for which the time-averaged driving force acts on the head and the counter-forces are distributed along the orientation vector axis to satisfy the force-free condition.
Note that the stresslet magnitude of the coarse-grained pusher is the time-averaged value of the full model.
The second model features a coarse-grained undulatory swimmer, similar to that of \cite{IshimotoPRL}, for which time-varying driving forces are applied to representative points (Figs.\ref{fig4}b, c).
The time-varying driving force is calculated by reference to the full model. We then integrate the hydrodynamic forces acting on sections of the flagellum; we divide the flagellum into three or ten parts.
Each spermatozoon is initially placed at random and the later swimming behaviour in suspension investigated (Fig. \ref{fig4}e).
Self-organisation is not achieved by the pusher in the absence of temporal variation in flagellar beating.
Undulatory swimmers with three-point forces evidence local collective swimming.
As the spatial resolution of the driving force increases, thus for swimmers with ten-point forces, coherent structures are observed.
The coarse-grained model neglects the fluid resistance of the flagellum, and the relaxation time is thus very short.
These results indicate that unsteady fluid interactions are important when coherent structures of active swimmers are observed, and suggest that the superimposition of time-averaged stress fields does not effectively track the collective structure of unsteady self-propelled particles.

{\it Conclusion--}
We developed a mechanical model of spermatozoal swimming; we explicitly expressed the elastohydrodynamic interactions among cells.
We confirmed that spermatozoa enter a polar order in bulk suspension and that this structure is stable because each spermatozoon becomes autonomously stable in an individual configuration.
Flagellar self-synchronisation via fluid motion affects the dynamics of flagellar beating, but the cell structure is less affected.
The comparison with a coarse-grained steady pusher highlights the importance of unsteady hydrodynamic interactions in terms of polar order formation, revealing a novel self-organising mechanism of active swimmers.
These results advance our knowledge on self-organisation of active matter and will contribute to analyses of biofluids and active fluids.

{\it Acknowledgments--}
The authors acknowledge the support of JSPS KAKENHI (21H04999, 21H05308) and JST PRESTO (JPMJPR2142).


\onecolumngrid

\clearpage

\clearpage

\clearpage

\clearpage
\begin{thebibliography}{41}
\expandafter\ifx\csname natexlab\endcsname\relax\def\natexlab#1{#1}\fi
\expandafter\ifx\csname bibnamefont\endcsname\relax
  \def\bibnamefont#1{#1}\fi
\expandafter\ifx\csname bibfnamefont\endcsname\relax
  \def\bibfnamefont#1{#1}\fi
\expandafter\ifx\csname citenamefont\endcsname\relax
  \def\citenamefont#1{#1}\fi
\expandafter\ifx\csname url\endcsname\relax
  \def\url#1{\texttt{#1}}\fi
\expandafter\ifx\csname urlprefix\endcsname\relax\def\urlprefix{URL }\fi
\providecommand{\bibinfo}[2]{#2}
\providecommand{\eprint}[2][]{\url{#2}}

\bibitem[{\citenamefont{Fauci and Dillon}(2006)}]{Fauci}
\bibinfo{author}{\bibfnamefont{L.~J.} \bibnamefont{Fauci}} \bibnamefont{and}
  \bibinfo{author}{\bibfnamefont{R.}~\bibnamefont{Dillon}},
  \bibinfo{journal}{Annual Review of Fluid Mechanics}
  \textbf{\bibinfo{volume}{38}}, \bibinfo{pages}{371} (\bibinfo{year}{2006}).

\bibitem[{\citenamefont{Gaffney et~al.}(2011)\citenamefont{Gaffney, Gadelha,
  Smith, Blake, and Kirkman-Brown}}]{Gaffney}
\bibinfo{author}{\bibfnamefont{E.}~\bibnamefont{Gaffney}},
  \bibinfo{author}{\bibfnamefont{H.}~\bibnamefont{Gadelha}},
  \bibinfo{author}{\bibfnamefont{D.}~\bibnamefont{Smith}},
  \bibinfo{author}{\bibfnamefont{J.}~\bibnamefont{Blake}}, \bibnamefont{and}
  \bibinfo{author}{\bibfnamefont{J.}~\bibnamefont{Kirkman-Brown}},
  \bibinfo{journal}{Annual Review of Fluid Mechanics}
  \textbf{\bibinfo{volume}{43}}, \bibinfo{pages}{501} (\bibinfo{year}{2011}).

\bibitem[{\citenamefont{Chakrabarti and Saintillan}(2019)}]{Chakrabarti}
\bibinfo{author}{\bibfnamefont{B.}~\bibnamefont{Chakrabarti}} \bibnamefont{and}
  \bibinfo{author}{\bibfnamefont{D.}~\bibnamefont{Saintillan}},
  \bibinfo{journal}{Physical Review Fluids} \textbf{\bibinfo{volume}{4}},
  \bibinfo{pages}{043102} (\bibinfo{year}{2019}).

\bibitem[{\citenamefont{Fauci and McDonald}(1995)}]{Fauci2}
\bibinfo{author}{\bibfnamefont{L.~J.} \bibnamefont{Fauci}} \bibnamefont{and}
  \bibinfo{author}{\bibfnamefont{A.}~\bibnamefont{McDonald}},
  \bibinfo{journal}{Bulletin of Mathematical Biology}
  \textbf{\bibinfo{volume}{57}}, \bibinfo{pages}{679 } (\bibinfo{year}{1995}),
  ISSN \bibinfo{issn}{0092-8240}.

\bibitem[{\citenamefont{Ishimoto et~al.}(2017)\citenamefont{Ishimoto,
  Gad\^elha, Gaffney, Smith, and Kirkman-Brown}}]{IshimotoPRL}
\bibinfo{author}{\bibfnamefont{K.}~\bibnamefont{Ishimoto}},
  \bibinfo{author}{\bibfnamefont{H.}~\bibnamefont{Gad\^elha}},
  \bibinfo{author}{\bibfnamefont{E.~A.} \bibnamefont{Gaffney}},
  \bibinfo{author}{\bibfnamefont{D.~J.} \bibnamefont{Smith}}, \bibnamefont{and}
  \bibinfo{author}{\bibfnamefont{J.}~\bibnamefont{Kirkman-Brown}},
  \bibinfo{journal}{Physical Review Letters} \textbf{\bibinfo{volume}{118}},
  \bibinfo{pages}{124501} (\bibinfo{year}{2017}).

\bibitem[{\citenamefont{Smith et~al.}(2009)\citenamefont{Smith, Gaffney,
  Gadelha, Kapur, and Kirkman-Brown}}]{Smith}
\bibinfo{author}{\bibfnamefont{D.~J.} \bibnamefont{Smith}},
  \bibinfo{author}{\bibfnamefont{E.~A.} \bibnamefont{Gaffney}},
  \bibinfo{author}{\bibfnamefont{H.}~\bibnamefont{Gadelha}},
  \bibinfo{author}{\bibfnamefont{N.}~\bibnamefont{Kapur}}, \bibnamefont{and}
  \bibinfo{author}{\bibfnamefont{J.~C.} \bibnamefont{Kirkman-Brown}},
  \bibinfo{journal}{Cell Motility and the Cytoskeleton}
  \textbf{\bibinfo{volume}{66}}, \bibinfo{pages}{220} (\bibinfo{year}{2009}).

\bibitem[{\citenamefont{Saintillan}(2018)}]{Saintillan}
\bibinfo{author}{\bibfnamefont{D.}~\bibnamefont{Saintillan}},
  \bibinfo{journal}{Annual Review of Fluid Mechanics}
  \textbf{\bibinfo{volume}{50}}, \bibinfo{pages}{563} (\bibinfo{year}{2018}).

\bibitem[{\citenamefont{Miles et~al.}(2019)\citenamefont{Miles, Evans, Shelley,
  and Spagnolie}}]{Miles}
\bibinfo{author}{\bibfnamefont{C.~J.} \bibnamefont{Miles}},
  \bibinfo{author}{\bibfnamefont{A.~A.} \bibnamefont{Evans}},
  \bibinfo{author}{\bibfnamefont{M.~J.} \bibnamefont{Shelley}},
  \bibnamefont{and} \bibinfo{author}{\bibfnamefont{S.~E.}
  \bibnamefont{Spagnolie}}, \bibinfo{journal}{Physical Review Letters}
  \textbf{\bibinfo{volume}{122}}, \bibinfo{pages}{098002}
  (\bibinfo{year}{2019}).

\bibitem[{\citenamefont{Ishikawa et~al.}(2022)\citenamefont{Ishikawa,
  Nghi~Dang, and Lauga}}]{Ishikawa}
\bibinfo{author}{\bibfnamefont{T.}~\bibnamefont{Ishikawa}},
  \bibinfo{author}{\bibfnamefont{T.}~\bibnamefont{Nghi~Dang}},
  \bibnamefont{and} \bibinfo{author}{\bibfnamefont{E.}~\bibnamefont{Lauga}},
  \bibinfo{journal}{Physical Review Fluids} \textbf{\bibinfo{volume}{7}},
  \bibinfo{pages}{013104} (\bibinfo{year}{2022}).

\bibitem[{\citenamefont{Creppy et~al.}(2015)\citenamefont{Creppy, Praud,
  Druart, Kohnke, and Plouraboue}}]{Creppy}
\bibinfo{author}{\bibfnamefont{A.}~\bibnamefont{Creppy}},
  \bibinfo{author}{\bibfnamefont{O.}~\bibnamefont{Praud}},
  \bibinfo{author}{\bibfnamefont{X.}~\bibnamefont{Druart}},
  \bibinfo{author}{\bibfnamefont{P.~L.} \bibnamefont{Kohnke}},
  \bibnamefont{and}
  \bibinfo{author}{\bibfnamefont{F.}~\bibnamefont{Plouraboue}},
  \bibinfo{journal}{Physical Review E} \textbf{\bibinfo{volume}{92}},
  \bibinfo{pages}{032722} (\bibinfo{year}{2015}).

\bibitem[{\citenamefont{Ishikawa et~al.}(2011)\citenamefont{Ishikawa, Yoshida,
  Ueno, Wideman, Imai, and Yamaguchi}}]{IshikawaPRL}
\bibinfo{author}{\bibfnamefont{T.}~\bibnamefont{Ishikawa}},
  \bibinfo{author}{\bibfnamefont{N.}~\bibnamefont{Yoshida}},
  \bibinfo{author}{\bibfnamefont{H.}~\bibnamefont{Ueno}},
  \bibinfo{author}{\bibfnamefont{M.}~\bibnamefont{Wideman}},
  \bibinfo{author}{\bibfnamefont{Y.}~\bibnamefont{Imai}}, \bibnamefont{and}
  \bibinfo{author}{\bibfnamefont{T.}~\bibnamefont{Yamaguchi}},
  \bibinfo{journal}{Physical Review Letters} \textbf{\bibinfo{volume}{107}},
  \bibinfo{pages}{028102} (\bibinfo{year}{2011}).

\bibitem[{\citenamefont{Ishimoto and Gaffney}(2018)}]{Ishimoto3}
\bibinfo{author}{\bibfnamefont{K.}~\bibnamefont{Ishimoto}} \bibnamefont{and}
  \bibinfo{author}{\bibfnamefont{E.~A.} \bibnamefont{Gaffney}},
  \bibinfo{journal}{Scientific Reports} \textbf{\bibinfo{volume}{8}},
  \bibinfo{pages}{15600} (\bibinfo{year}{2018}).

\bibitem[{\citenamefont{Schoeller and Keaveny}(2018)}]{Schoeller}
\bibinfo{author}{\bibfnamefont{S.~F.} \bibnamefont{Schoeller}}
  \bibnamefont{and} \bibinfo{author}{\bibfnamefont{E.~E.}
  \bibnamefont{Keaveny}}, \bibinfo{journal}{Journal of Royal Society
  Interface.} \textbf{\bibinfo{volume}{15}}, \bibinfo{pages}{20170834}
  (\bibinfo{year}{2018}).

\bibitem[{\citenamefont{Tung et~al.}(2017)\citenamefont{Tung, Lin, Harvey,
  Fiore, Ardon, Wu, and Suarez}}]{Tung}
\bibinfo{author}{\bibfnamefont{C.}~\bibnamefont{Tung}},
  \bibinfo{author}{\bibfnamefont{C.}~\bibnamefont{Lin}},
  \bibinfo{author}{\bibfnamefont{B.}~\bibnamefont{Harvey}},
  \bibinfo{author}{\bibfnamefont{A.~G.} \bibnamefont{Fiore}},
  \bibinfo{author}{\bibfnamefont{F.}~\bibnamefont{Ardon}},
  \bibinfo{author}{\bibfnamefont{M.}~\bibnamefont{Wu}}, \bibnamefont{and}
  \bibinfo{author}{\bibfnamefont{S.~S.} \bibnamefont{Suarez}},
  \bibinfo{journal}{Scientific Reports} \textbf{\bibinfo{volume}{7}},
  \bibinfo{pages}{3152} (\bibinfo{year}{2017}).

\bibitem[{\citenamefont{E. et~al.}(2014)\citenamefont{E., Wioland, and
  Goldstein}}]{Lushi}
\bibinfo{author}{\bibfnamefont{L.}~\bibnamefont{E.}},
  \bibinfo{author}{\bibfnamefont{H.}~\bibnamefont{Wioland}}, \bibnamefont{and}
  \bibinfo{author}{\bibfnamefont{R.~E.} \bibnamefont{Goldstein}},
  \bibinfo{journal}{Proceedings of the National Academy of Sciences of the
  United States of America} \textbf{\bibinfo{volume}{111}},
  \bibinfo{pages}{9733} (\bibinfo{year}{2014}).

\bibitem[{\citenamefont{Omori et~al.}(2022)\citenamefont{Omori, Kikuchi,
  Schmitz, Pavlovic, Chuang, and Ishikawa}}]{OmoriJFM}
\bibinfo{author}{\bibfnamefont{T.}~\bibnamefont{Omori}},
  \bibinfo{author}{\bibfnamefont{K.}~\bibnamefont{Kikuchi}},
  \bibinfo{author}{\bibfnamefont{M.}~\bibnamefont{Schmitz}},
  \bibinfo{author}{\bibfnamefont{M.}~\bibnamefont{Pavlovic}},
  \bibinfo{author}{\bibfnamefont{C.-H.} \bibnamefont{Chuang}},
  \bibnamefont{and} \bibinfo{author}{\bibfnamefont{T.}~\bibnamefont{Ishikawa}},
  \bibinfo{journal}{Journal of Fluid Mechanics} \textbf{\bibinfo{volume}{930}},
  \bibinfo{pages}{A30} (\bibinfo{year}{2022}).

\bibitem[{\citenamefont{Samatas and Lintuvuori}(2023)}]{Samatas}
\bibinfo{author}{\bibfnamefont{S.}~\bibnamefont{Samatas}} \bibnamefont{and}
  \bibinfo{author}{\bibfnamefont{J.}~\bibnamefont{Lintuvuori}},
  \bibinfo{journal}{Physical Review Letters} \textbf{\bibinfo{volume}{130}},
  \bibinfo{pages}{024001} (\bibinfo{year}{2023}).

\bibitem[{\citenamefont{Omori and Ishikawa}(2019)}]{OmoriMaxwell}
\bibinfo{author}{\bibfnamefont{T.}~\bibnamefont{Omori}} \bibnamefont{and}
  \bibinfo{author}{\bibfnamefont{T.}~\bibnamefont{Ishikawa}},
  \bibinfo{journal}{Micromachine} \textbf{\bibinfo{volume}{10}},
  \bibinfo{pages}{78} (\bibinfo{year}{2019}).

\bibitem[{\citenamefont{Rikmenspoel}(1984)}]{Rikmenspoel}
\bibinfo{author}{\bibfnamefont{R.}~\bibnamefont{Rikmenspoel}},
  \bibinfo{journal}{Journal of Experimental Biology}
  \textbf{\bibinfo{volume}{108}}, \bibinfo{pages}{205} (\bibinfo{year}{1984}).

\bibitem[{\citenamefont{Camalet and Julicher}(2000)}]{Camalet}
\bibinfo{author}{\bibfnamefont{S.}~\bibnamefont{Camalet}} \bibnamefont{and}
  \bibinfo{author}{\bibfnamefont{F.}~\bibnamefont{Julicher}},
  \bibinfo{journal}{New Journal of Physics} \textbf{\bibinfo{volume}{2}},
  \bibinfo{pages}{24.1} (\bibinfo{year}{2000}).

\bibitem[{\citenamefont{Chakrabarti and Sintillan}(2019)}]{ChakrabartiPRL}
\bibinfo{author}{\bibfnamefont{B.}~\bibnamefont{Chakrabarti}} \bibnamefont{and}
  \bibinfo{author}{\bibfnamefont{D.}~\bibnamefont{Sintillan}},
  \bibinfo{journal}{Physical Review Letters} \textbf{\bibinfo{volume}{123}},
  \bibinfo{pages}{208101} (\bibinfo{year}{2019}).

\bibitem[{\citenamefont{Goldstein et~al.}(2016)\citenamefont{Goldstein, Lauga,
  Pesci, and Proctor}}]{Goldstein}
\bibinfo{author}{\bibfnamefont{R.~E.} \bibnamefont{Goldstein}},
  \bibinfo{author}{\bibfnamefont{E.}~\bibnamefont{Lauga}},
  \bibinfo{author}{\bibfnamefont{A.~I.} \bibnamefont{Pesci}}, \bibnamefont{and}
  \bibinfo{author}{\bibfnamefont{M.~R.~E.} \bibnamefont{Proctor}},
  \bibinfo{journal}{Physical Review Fluids} \textbf{\bibinfo{volume}{1}},
  \bibinfo{pages}{073201} (\bibinfo{year}{2016}).

\bibitem[{\citenamefont{Oriola et~al.}(2017)\citenamefont{Oriola, Gadelha, and
  Casademunt}}]{Oriola}
\bibinfo{author}{\bibfnamefont{D.}~\bibnamefont{Oriola}},
  \bibinfo{author}{\bibfnamefont{H.}~\bibnamefont{Gadelha}}, \bibnamefont{and}
  \bibinfo{author}{\bibfnamefont{J.}~\bibnamefont{Casademunt}},
  \bibinfo{journal}{Royal Society Open Science} \textbf{\bibinfo{volume}{4}},
  \bibinfo{pages}{160698} (\bibinfo{year}{2017}).

\bibitem[{\citenamefont{Riedel-Kruse et~al.}(2007)\citenamefont{Riedel-Kruse,
  Hilfinger, Howard, and Julicher}}]{Riedel-Kruse}
\bibinfo{author}{\bibfnamefont{I.}~\bibnamefont{Riedel-Kruse}},
  \bibinfo{author}{\bibfnamefont{A.}~\bibnamefont{Hilfinger}},
  \bibinfo{author}{\bibfnamefont{J.}~\bibnamefont{Howard}}, \bibnamefont{and}
  \bibinfo{author}{\bibfnamefont{F.}~\bibnamefont{Julicher}},
  \bibinfo{journal}{HFSP Journal} \textbf{\bibinfo{volume}{1}},
  \bibinfo{pages}{192} (\bibinfo{year}{2007}).

\bibitem[{\citenamefont{Sartori et~al.}(2016)\citenamefont{Sartori, Geyer,
  Scholich, Julicher, and Howard}}]{Sartori}
\bibinfo{author}{\bibfnamefont{P.}~\bibnamefont{Sartori}},
  \bibinfo{author}{\bibfnamefont{V.~F.} \bibnamefont{Geyer}},
  \bibinfo{author}{\bibfnamefont{A.}~\bibnamefont{Scholich}},
  \bibinfo{author}{\bibfnamefont{F.}~\bibnamefont{Julicher}}, \bibnamefont{and}
  \bibinfo{author}{\bibfnamefont{J.}~\bibnamefont{Howard}},
  \bibinfo{journal}{eLife.} \textbf{\bibinfo{volume}{5}},
  \bibinfo{pages}{e13258} (\bibinfo{year}{2016}).

\bibitem[{\citenamefont{Golestanian et~al.}(2011)\citenamefont{Golestanian,
  Yeomans, and Uchida}}]{Ramin}
\bibinfo{author}{\bibfnamefont{R.}~\bibnamefont{Golestanian}},
  \bibinfo{author}{\bibfnamefont{J.~M.} \bibnamefont{Yeomans}},
  \bibnamefont{and} \bibinfo{author}{\bibfnamefont{N.}~\bibnamefont{Uchida}},
  \bibinfo{journal}{Soft Matter} \textbf{\bibinfo{volume}{7}},
  \bibinfo{pages}{3074} (\bibinfo{year}{2011}).

\bibitem[{\citenamefont{Miki and Clapham}(2013)}]{Miki}
\bibinfo{author}{\bibfnamefont{K.}~\bibnamefont{Miki}} \bibnamefont{and}
  \bibinfo{author}{\bibfnamefont{D.~E.} \bibnamefont{Clapham}},
  \bibinfo{journal}{Current Biology} \textbf{\bibinfo{volume}{23}},
  \bibinfo{pages}{443} (\bibinfo{year}{2013}).

\bibitem[{\citenamefont{Kantsler et~al.}(2014)\citenamefont{Kantsler, Dunkel,
  Blayney, and Goldstein}}]{Kantsler}
\bibinfo{author}{\bibfnamefont{V.}~\bibnamefont{Kantsler}},
  \bibinfo{author}{\bibfnamefont{J.}~\bibnamefont{Dunkel}},
  \bibinfo{author}{\bibfnamefont{M.}~\bibnamefont{Blayney}}, \bibnamefont{and}
  \bibinfo{author}{\bibfnamefont{R.~E.} \bibnamefont{Goldstein}},
  \bibinfo{journal}{eLife} \textbf{\bibinfo{volume}{3}},
  \bibinfo{pages}{e02403} (\bibinfo{year}{2014}).

\bibitem[{\citenamefont{Ishimoto and Gaffney}(2015)}]{Ishimoto2}
\bibinfo{author}{\bibfnamefont{K.}~\bibnamefont{Ishimoto}} \bibnamefont{and}
  \bibinfo{author}{\bibfnamefont{E.~A.} \bibnamefont{Gaffney}},
  \bibinfo{journal}{Journal of The Royal Society Interface}
  \textbf{\bibinfo{volume}{12}}, \bibinfo{pages}{20150172}
  (\bibinfo{year}{2015}).

\bibitem[{\citenamefont{Omori and Ishikawa}(2016)}]{OmoriPRE}
\bibinfo{author}{\bibfnamefont{T.}~\bibnamefont{Omori}} \bibnamefont{and}
  \bibinfo{author}{\bibfnamefont{T.}~\bibnamefont{Ishikawa}},
  \bibinfo{journal}{Physical Review E} \textbf{\bibinfo{volume}{93}},
  \bibinfo{pages}{032402} (\bibinfo{year}{2016}).

\bibitem[{\citenamefont{Ito et~al.}(2019)\citenamefont{Ito, Omori, and
  Ishikawa}}]{Ito}
\bibinfo{author}{\bibfnamefont{H.}~\bibnamefont{Ito}},
  \bibinfo{author}{\bibfnamefont{T.}~\bibnamefont{Omori}}, \bibnamefont{and}
  \bibinfo{author}{\bibfnamefont{T.}~\bibnamefont{Ishikawa}},
  \bibinfo{journal}{J. Fluid Mech.} \textbf{\bibinfo{volume}{874}},
  \bibinfo{pages}{774} (\bibinfo{year}{2019}).

\bibitem[{\citenamefont{Omori et~al.}(2020)\citenamefont{Omori, Ito, and
  Ishikawa}}]{OmoriPNAS}
\bibinfo{author}{\bibfnamefont{T.}~\bibnamefont{Omori}},
  \bibinfo{author}{\bibfnamefont{H.}~\bibnamefont{Ito}}, \bibnamefont{and}
  \bibinfo{author}{\bibfnamefont{T.}~\bibnamefont{Ishikawa}},
  \bibinfo{journal}{Proceedings of the National Academy of Sciences of the
  United States of America} \textbf{\bibinfo{volume}{117}},
  \bibinfo{pages}{30201} (\bibinfo{year}{2020}).

\bibitem[{\citenamefont{Nanami et~al.}(2020)\citenamefont{Nanami, Omori, and
  Ishikawa}}]{Taketoshi}
\bibinfo{author}{\bibfnamefont{T.}~\bibnamefont{Nanami}},
  \bibinfo{author}{\bibfnamefont{T.}~\bibnamefont{Omori}}, \bibnamefont{and}
  \bibinfo{author}{\bibfnamefont{T.}~\bibnamefont{Ishikawa}},
  \bibinfo{journal}{Physics of Fluids} \textbf{\bibinfo{volume}{32}},
  \bibinfo{pages}{101901} (\bibinfo{year}{2020}).

\bibitem[{\citenamefont{Tornberg and Shelley}(2004)}]{Tornberg}
\bibinfo{author}{\bibfnamefont{A.-K.} \bibnamefont{Tornberg}} \bibnamefont{and}
  \bibinfo{author}{\bibfnamefont{M.~J.} \bibnamefont{Shelley}},
  \bibinfo{journal}{Journal of Computational Physics}
  \textbf{\bibinfo{volume}{196}}, \bibinfo{pages}{8 } (\bibinfo{year}{2004}).

\bibitem[{\citenamefont{Beenakker}(1986)}]{Beenakker}
\bibinfo{author}{\bibfnamefont{C.~W.~J.} \bibnamefont{Beenakker}},
  \bibinfo{journal}{Journal of Chemical Physics} \textbf{\bibinfo{volume}{85}},
  \bibinfo{pages}{1581} (\bibinfo{year}{1986}).

\bibitem[{\citenamefont{Evans et~al.}(2011)\citenamefont{Evans, Ishikawa,
  Yamaguchi, and Lauga}}]{Evans}
\bibinfo{author}{\bibfnamefont{A.~A.} \bibnamefont{Evans}},
  \bibinfo{author}{\bibfnamefont{T.}~\bibnamefont{Ishikawa}},
  \bibinfo{author}{\bibfnamefont{T.}~\bibnamefont{Yamaguchi}},
  \bibnamefont{and} \bibinfo{author}{\bibfnamefont{E.}~\bibnamefont{Lauga}},
  \bibinfo{journal}{Physics of Fluids} \textbf{\bibinfo{volume}{23}},
  \bibinfo{pages}{111702} (\bibinfo{year}{2011}).

\bibitem[{\citenamefont{Alert et~al.}(2022)\citenamefont{Alert, Casademunt, and
  Joanny}}]{Alert}
\bibinfo{author}{\bibfnamefont{R.}~\bibnamefont{Alert}},
  \bibinfo{author}{\bibfnamefont{J.}~\bibnamefont{Casademunt}},
  \bibnamefont{and} \bibinfo{author}{\bibfnamefont{J.-F.}
  \bibnamefont{Joanny}}, \bibinfo{journal}{Annual Review of Condensed Matter
  Physics} \textbf{\bibinfo{volume}{13}}, \bibinfo{pages}{143}
  (\bibinfo{year}{2022}).

\bibitem[{\citenamefont{Ramaswamy}(2019)}]{Ramaswamy}
\bibinfo{author}{\bibfnamefont{S.}~\bibnamefont{Ramaswamy}},
  \bibinfo{journal}{Nature Reviews Physics} \textbf{\bibinfo{volume}{1}},
  \bibinfo{pages}{640} (\bibinfo{year}{2019}).

\bibitem[{\citenamefont{Ishikawa}(2009)}]{Ishikawa2009}
\bibinfo{author}{\bibfnamefont{T.}~\bibnamefont{Ishikawa}},
  \bibinfo{journal}{Journal of the Royal Society Interface}
  \textbf{\bibinfo{volume}{6}}, \bibinfo{pages}{815} (\bibinfo{year}{2009}).

\bibitem[{\citenamefont{Ishikawa et~al.}(2020)\citenamefont{Ishikawa, Omori,
  and Kikuchi}}]{IshikawaAPL}
\bibinfo{author}{\bibfnamefont{T.}~\bibnamefont{Ishikawa}},
  \bibinfo{author}{\bibfnamefont{T.}~\bibnamefont{Omori}}, \bibnamefont{and}
  \bibinfo{author}{\bibfnamefont{K.}~\bibnamefont{Kikuchi}},
  \bibinfo{journal}{APL Bioengineering} \textbf{\bibinfo{volume}{4}},
  \bibinfo{pages}{041504} (\bibinfo{year}{2020}).

\bibitem[{\citenamefont{Lindemann}(1994)}]{Lindemann}
\bibinfo{author}{\bibfnamefont{C.~B.} \bibnamefont{Lindemann}},
  \bibinfo{journal}{Journal of Theoretical Biology}
  \textbf{\bibinfo{volume}{168}}, \bibinfo{pages}{175} (\bibinfo{year}{1994}).

\end{thebibliography}
\end{document}


\preprint{AAPM/123-QED}

\title{Self-organisation of spermatozoa in bulk suspensions\\via unsteady elastohydrodynamic interactions}

\author{Nanami Taketoshi\footnote{Address: Aramaki Aoba 6-6-01, Sendai, Miyagi, Japan}}
\email{nanami.taketoshi.q1@dc.tohoku.ac.jp}

\author{Toshihiro Omori}
\email{omori@tohoku.ac.jp}

\author{Takuji Ishikawa}
\affiliation{Department of Biomedical Engineering, Tohoku University}
\affiliation{Department of Finemechanics, Tohoku University}
\date{\today}

\maketitle
In this document, we present governing equations, numerical methods, and additional results linked to the main text.
Details of our methodology can be found in out previous article[32].

\section {Governing Equations and Numerical Methods}
\subsection {Fluid mechanics of swimming sperm}
We applied slender-body theory [32]
to describe flagellar dynamics, since a flagellum is very thin compared to its length.
The flow field at point ${\bm x}$ located on the $i$-th sperm's flagellum, ${\bm x} \in s_i$, is given by [30]:
\begin{eqnarray}
\label{fla_v}
{\bm v}({\bm x}) = &-& \frac{1}{8\pi\mu}{\bm \Lambda}({\bm x}) \cdot {\bm q}_f({\bm x}) \nonumber\\
&-&\frac{1}{8\pi\mu}\int_{s_i}({\bm J}({\bm x}, {\bm y})\cdot {\bm q}_f({\bm y})-{\bm K}({\bm x}, {\bm y})\cdot {\bm q}_f({\bm x}))ds({\bm y})\nonumber\\
&-& \frac{1}{8\pi\mu}\sum_{j \neq i}^N \int_{s_j}[{\bm J}({\bm x}, {\bm y})+{\bm W}({\bm x}, {\bm y})]\cdot {\bm q}_f({\bm y})ds({\bm y})\nonumber\\
&+&\frac{1}{8\pi\mu} \sum_{j=1}^N{\bm J}({\bm x},{\bm y}_j)\cdot {\bm F}^{h}_{j}({\bm y}_j)
\end{eqnarray}
where ${\bm q}_f$ is the force density exerted on the flagellum, and ${\bm F}^h$ is the force on the head.
${\bm J}$ is the Green's function, given by:
\begin{eqnarray}
\label{2_eq2}
J_{ij} = \frac{\delta_{ij}}{r} + \frac{r_ir_j}{r^3},
\end{eqnarray}
where $\delta_{ij}$ is the Kronecker's delta$, r = |{\bm r}|$, and ${\bm r} = {\bm x} - {\bm y}$. ${\bm \Lambda}$, ${\bm K}$, and ${\bm W}$ are the operators on the slender body, which are given by:
\begin{eqnarray}
\label{2_eq3}
{\Lambda_{ij}}({\bm x}) &=& c(\delta_{ij}+t_i({\bm x}) t_j({\bm x})) + 2(\delta_{ij} - t_i({\bm x}) t_j({\bm x})),\\
\label{2_eq4}
{K_{ij}}({\bm x},{\bm y}) &=& -\frac{\delta_{ij} + t_i({\bm x}) t_j({\bm x})}{|s({\bm x})-s({\bm y})|}, \\
\label{2_eq5}
{W_{ij}}({\bm x},{\bm y}) &=& \frac{(\varepsilon L)^2}{2}
\left[ \frac{\delta_{ij}}{r^3} - 3\frac{r_i r_j}{r^5} \right],
\end{eqnarray}
where $c = -\ln(\varepsilon^2e)$, 
$\varepsilon = a_{fla}/L$, and $a_{fla}$ is the flagellar radius.
To mimic the configuration of human sperm [2], $\varepsilon$ is set to $\varepsilon = 2.3 \times 10^{-3}$.
\\
\ The sperm head is assumed to be a sphere of radius $a$ ($a/L = 0.042)$, and its drag is determined by Stokes' law [29]. 
When an observation point ${\bm x}$ is located on the $i$-th sperm head, the flow field is given in a similar manner:
\begin{eqnarray}
\label{head_v}
{\bm v}({\bm x}) &=&\frac{{\bm F}^{h}_{i}}{6\pi\mu a}  \nonumber \\
&-&\frac{1}{8\pi\mu}\sum_{j}^N \int_{s_j}[{\bm J}({\bm x}, {\bm y})+{\bm W}({\bm x}, {\bm y})]\cdot {\bm q}_f({\bm y})ds_j({\bm y})\nonumber\\
&+&\frac{1}{8\pi\mu} \sum_{j \neq i}^N{\bm J}({\bm x},{\bm y}_j)\cdot {\bm F}^{h}_{j}({\bm y}_j) .
\end{eqnarray}

\ To express force-free of sperm cells, ${\bm F}^h$ and ${\bm q}_f$ have the following relationship:
\begin{eqnarray}
\label{F_int}
 {\bm F}_i^h = \int_{s_i}{\bm q}_{f}ds,
 \end{eqnarray}

\ In the case of human or bull sperm, the neck behaves as a rigid connection [2]. The head-tail junction is thus assumed as a rigid connection, and the boundary condition is given as:
\begin{eqnarray}
{\bm v}({\bm x}_{h}) = {\bm v}({\bm x}_0) = {\bm v}({\bm x}_{neck}),
 \end{eqnarray}
where ${\bm x}_{h}$, ${\bm x}_{0}$, and ${\bm x}_{neck}$ are the position of the head, head-tail junction, and base of the flagellum, respectively.

\subsection {Solid mechanics}
\ To express fluid-structure interactions, the solid mechanics of a beating flagellum are also taken into account. Assume that the flagellum is an isotropic elastic material. Due to its slenderness and the small wave number of typical flagellar deformations, the flagellum is modeled by an Euler-Bernoulli rod, and we neglect the inner shear stress of the flagellum.
Elastic energy per unit length, $W$, is given by [20]:
\begin{eqnarray}
\label{W}
W&=&\frac{1}{2}E_bC^2  + \frac{1}{2}\tau \delta s^2,
\end{eqnarray}
where $E_b$ is the bending rigidity, and $C$ is the curvature.
The last term is the Lagrange multiplier, $\tau$, acting on the flagellum to resist extension or compression, $\delta s$. Based on the principal of virtual work, the weak form of the equilibrium equation is given by:
\begin{eqnarray}
\label{W2}
\int\hat{\bm u}\cdot {\bm q}_{e} ds &=& \int\Delta W ds, \\
\label{W3}
\Delta W &=& E_bC\delta C + \tau \delta s,
\end{eqnarray}
where $\hat{\bm u}$ is the virtual displacement, $\delta C$ is the increment in the curvature by virtual displacement, and ${\bm q}_{ e}$ is the force density due to the elastic energy.
The Lagrange multiplier, $\tau$, is determined by an iterative method 
to preserve the flagellum length ($\delta s/L \le 0.01$).

\subsection {Active dynein force}
\ The internal motor protein dynein generates a sliding force along the flagellum.
To express the internal active force, we assume a balance among the hydrodynamic force ${\bm q}_{f}$, the elastic force ${\bm q}_{e}$, and the driving force ${\bm q}_{a}$:
\begin{eqnarray}
\label{2_eq13}
{\bm q}_{f} + {\bm q}_{e} + {\bm q}_{a} = 0.
\end{eqnarray}
We calculated the driving force ${\bm q}_{a}$ separately from the main calculation using a prescribed waveform observed experimentally [32]. 
Once ${\bm q}_{a}$ is given as a function of time and position, the driving force is treated as the input source in the main simulation.
During the main simulation, the flagellar waveform can be varied by the elastohydrodynamic interactions.

The prescribed waveform for ${\bm q}_{a}$ setting is given by the following procedure.
An orthonormal body frame is defined as $\xi_i$.
The body frame origin is equivalent to ${\bm x}_{neck}$.
The time-dependent reference waveform is described by the following simple formula [29,32].
\begin{eqnarray}
\label{eq:wave}
\xi_2 &=& A\xi_1\cos(k\xi_1/L - 2\pi f_0 t)\nonumber\\
\xi_3 &=& -\alpha A\xi_1\sin(k\xi_1/L - 2\pi f_0 t),
\end{eqnarray}
where $A (=0.018L)$ is the amplitude of the beat, $k (= 2\pi)$ is the wave number, $f_0$ is the frequency, and $\alpha (= 0.2)$ is the helicity parameter.
Once ${\bm q}_{a}$ is given, the fluid-structure interactions among sperm suspension can be solved by Eqs.(\ref{fla_v}) and (\ref{W2}).
Thus, the flagellar beat pattern is not prescribed in the main simulation. 
\subsection{Application of geometric clutch model}
\begin{figure}[h]
\begin{center}
\resizebox{0.6\textwidth}{!}{
\includegraphics[]{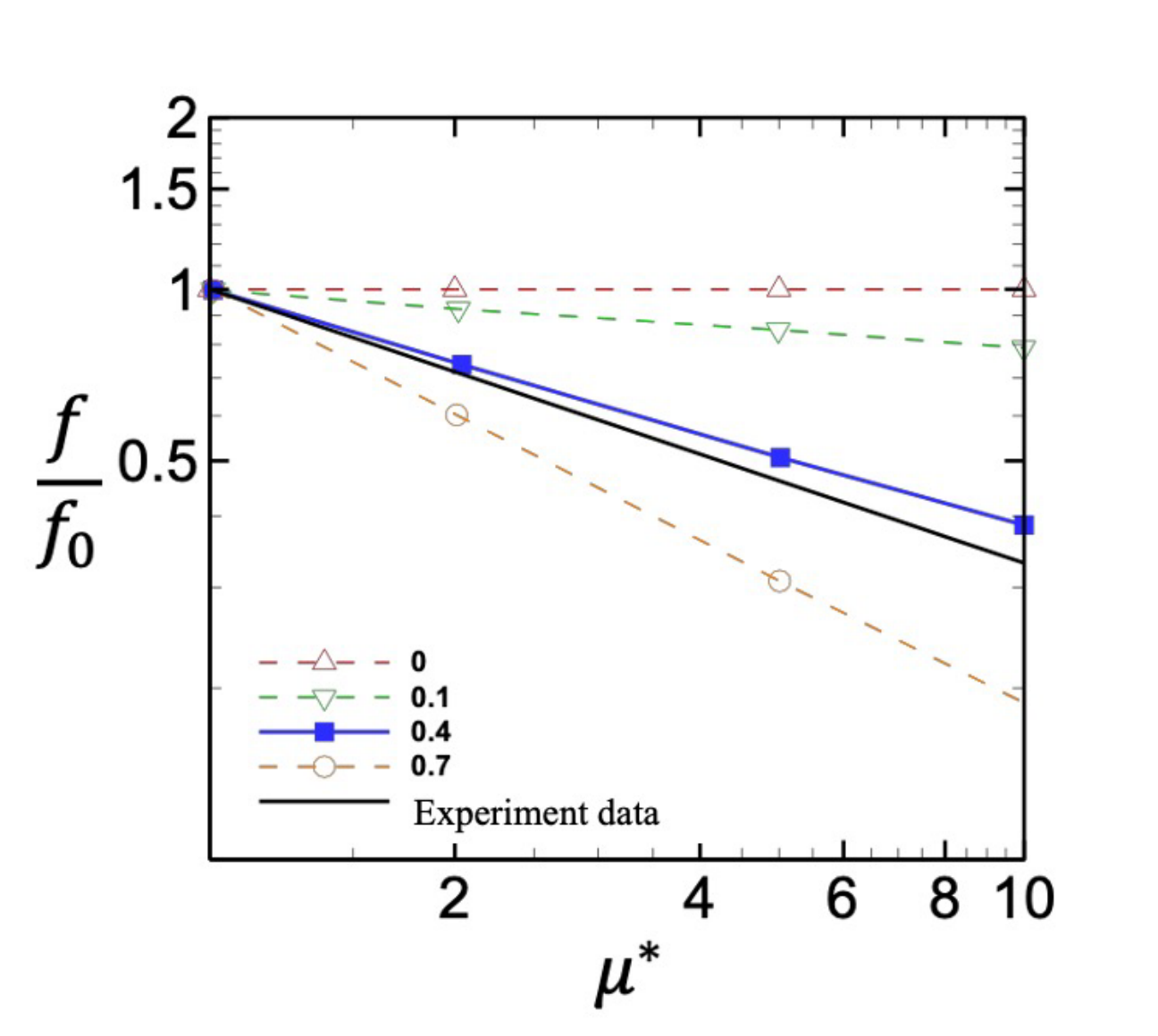}}
\caption{Beat frequency as a function of normalized viscosity. Each symbol represents different exponential indices.}
\label{F_er}
\end{center}
\end{figure}

We applied geometric clutch model to represent a curvature-dependent active force propagation [32].
The frequency of the active force is described by the following equation as a function of the time variation of the mean curvature:
\begin{equation}
\label{eq:freq}
\frac{f}{f_0} = \left(
\frac{1}{\dot{C_0}}\frac{\partial C_{ave}(t)}{\partial t}
\right)^{d},
\end{equation}
and
\begin{equation}
C_{ave}(t) = \frac{1}{L}\int_0^L C(s,t)ds
~,~~~
\dot{C_0} = \frac{1}{T} \int_0^T \frac{\partial C_{ave,0}(t)}{\partial t} dt ,
\end{equation}
where $f$ is the current frequency, $C(s,t)$ is the local curvature, and $C_{ave}(t)$ is the length-averaged curvature. 
The subscript 0 indicates the reference values, which are estimated from single swimming.
We set $d=0.4$ in order to coincide with [19] (cf. Fig.\ref{F_er}).
 \subsection {Ewald summation}
 \ To simulate sperm swimming in a bulk suspension, we applied triply periodic boundary conditions with Ewald summation.
Set up a cubic domain with one side $L_c = 4.0L$, and applied the infinite lattice sums:
\begin{eqnarray}
{\bm v}({\bm x})&\sim& {\bm v}^{self}({\bm x}) - \frac{1}{8\pi\mu}\sum_{\gamma}^{\infty}\sum_{j=1}^{N}\left\{\int_{s^{\gamma}_i}{\bm J}({\bm x}, {\bm y^{\gamma}})\cdot {\bm q}_f({\bm y^{\gamma}})ds_i({\bm y}) + {\bm J}({\bm x}, {\bm y^{\gamma}})\cdot {\bm F}^{h}_{j}({\bm y}^{\gamma})\right\}\nonumber \\
&\sim&  {\bm v}^{self}({\bm x}) - \frac{1}{8\pi\mu}\sum_{j=1}^{N}\left\{\int_{s^0_i}{\bm J^E}({\bm x}, {\bm y})\cdot {\bm q}_f({\bm y^0})ds_i({\bm y^0})+ {\bm J^E}({\bm x}, {\bm y^0})\cdot {\bm F}^{h}_{j}({\bm y}^0)\right\}
\end{eqnarray}
where $N$ is the number of sperm model in the unit domain, and $\gamma$ is the label of the periodic lattices. 
Since the self-terms $\bm{\Lambda}$ and ${\bm K}$ do not appear on periodic lattices and $\bm{W}$ decays faster than $\bm{J}$, we therefore consider the change of $\bm{J}$ in periodic boundary condition.
Green function ${\bm J}$ in periodic boundary condition is can be expressed as
\begin{eqnarray}
{\bm J^E}({\bm x}, {\bm y})=\sum_{\gamma}^{\infty}{\bm J}({\bm x}, {\bm y}^{\gamma}).
\end{eqnarray}
In this equation, $\gamma = 0$ means the original unit domain. 
When a point force is located at the point $\bm{x}_0$ in the original unit domain, 
the mirror point force in other domain $\gamma > 0$ is located at $\bm{x}_0+\bm{X}_{\gamma}$, as $\bm{X}_{\gamma} =(i_1L_x, i_2L_y,  i_3L_z)$, and $i_1, i_2, i_3$ are three integers.

\ In order to compute the ${\bm J}^E$ efficiently, we divided it as:
\begin{eqnarray}
\label{divide2}
&&\int{\bm J^E}(\bm{r})\cdot {\bm q}_f({\bm y})ds_i({\bm y})\nonumber\\
&=&\int \bm{J}^{E_I}(\bm{r})\cdot {\bm q}_f({\bm y})ds_i({\bm y})+\int \bm{J}^{E_{II}}(\bm{r})\cdot {\bm q}_f({\bm y})ds_i({\bm y})
\end{eqnarray}

\ From previous articles, Green function $J_{ij}$ can be expressed by Ewald summation as below:
 \begin{eqnarray}
J_{ij}^E(r)&=&\sum_{\gamma}J_{ij}^{E_I}({r_{\gamma}})+\frac{8\pi}{V}\sum_{\gamma}J_{ij}^{E_{II}}(k_\lambda)\\
J^{E_I}_{ij}(r) &=& \frac{\delta_{ij}}{k^2}E_1+\frac{r_ir_j}{r^3}E_2\\
J^{E_{II}}_{ij}(k) &=& (\frac{\delta_{ij}}{k^2}-\frac{k_ik_j}{k^4})(1+\frac{k^2}{4\xi^2}+\frac{k^4}{8\xi^4})\mathrm{exp}(-\frac{k^2}{4\xi^2})\mathrm{cos}(\bm{k}\cdot\bm{r}),
\end{eqnarray}
and the equation of flagellar velocity can be expressed as:
\begin{eqnarray}
\label{ewald_fla_v}
{\bm v}({\bm x}) = &-& \frac{1}{8\pi\mu}{\bm \Lambda}({\bm x}) \cdot {\bm q}_f({\bm x}) \nonumber \\
&-&\frac{1}{8\pi\mu}\sum_{j=1}^{N}\int_{s_i}({\bm J^E}({\bm x}, {\bm y})\cdot {\bm q}_f({\bm y})-{\bm K}({\bm x}, {\bm y})\cdot {\bm q}_f({\bm x}))ds_i({\bm y})\nonumber\\ 
&-&\frac{1}{8\pi\mu}\sum_{j \neq i}^N \int_{s_j}[{\bm J^E}({\bm x}, {\bm y})+{\bm W}({\bm x}, {\bm y})]\cdot {\bm q}_f({\bm y})ds_j({\bm y})\nonumber\\
&+&\frac{1}{8\pi\mu}\sum_{j=1}^N{\bm J^E}({\bm x},{\bm y}_j)\cdot {\bm F}^{h}_{j}({\bm y}_j).
\end{eqnarray}

Similarly to Eq.(\ref{ewald_fla_v}),the velocity of the head is also calculated is calculated by this equation.
\begin{eqnarray}
\label{ewald_v_head}
{\bm v}({\bm x}) = &-& \frac{1}{8\pi\mu}{\bm \Lambda}({\bm x}) \cdot {\bm q}_f({\bm x})\nonumber\\
&-&\sum_{1}^{N}\frac{1}{8\pi\mu}\int_{s_i}({\bm J^E}({\bm x}, {\bm y})\cdot {\bm q}_f({\bm y})-{\bm K}({\bm x}, {\bm y})\cdot {\bm q}_f({\bm x}))ds_i({\bm y}) \nonumber\\
&-&\frac{1}{8\pi\mu}\sum_{j \neq i}^N \int_{s_j}[{\bm J^E}({\bm x}, {\bm y})+{\bm W}({\bm x}, {\bm y})]\cdot {\bm q}_f({\bm y})ds_j({\bm y}) \nonumber\\
&+&\frac{1}{8\pi\mu}\sum_{j=1}^N{\bm J^E}({\bm x},{\bm y}_j)\cdot {\bm F}^{h}_{j}({\bm y}_j)+\frac{{\bm F^{h}_{j}}}{6\pi\mu a}.
\end{eqnarray}

\subsection {Multipole Expansion}
To accelerate the computation while maintaining accuracy, the multipole expansion method 
 is introduced in this study.
At the far field ($r >>1$), Taylor series expansion can be applied to the Green's function as follow:
\begin{eqnarray}
\label{J_m}
J_{ij}(\bm{x}-\bm{y}) &=& J_{ij}(\bm{x}-\bm{y}){\rvert}_{\bm{y} = \bm{y}^{\alpha}} + \frac{1}{1!}\frac{\partial J_{ij}(\bm{x}-\bm{y})}{\partial y_k}{\rvert}_{\bm{y} = \bm{y}^{\alpha}}(y_k-y_k^{\alpha})+...\nonumber\\
&=& J_{ij}(\bm{x}-\bm{y}^{\alpha}) - \frac{1}{1!}\frac{\partial J_{ij}(\bm{x}-\bm{y})}{\partial r_k}{\rvert}_{\bm{y} = \bm{y}^{\alpha}}\hat{r}_k^{\alpha}+...\nonumber\\
&=& \sum^{\infty}_{n=0}\left\{ \frac{(-1)^n}{n!}\frac{\partial^nJ_{ij}}{\partial r_k^n}{\rvert}_{\bm{y} = \bm{y}^{\alpha}}(\hat{r}_k^{\alpha})^n\right\},
\end{eqnarray}
where ${\bm y}^{\alpha}$ is the mass center of $\alpha$-th sperm, $\hat{r}_k^{\alpha} = y_k-y_k^{\alpha}$.
The boundary integral equation is then expressed as a superposition of near and far fields.
\begin{eqnarray}
v_i(\bm{x}) = v^{near}_i -\frac{1}{8\pi\mu}\sum^{N^{far}}_{\alpha}\sum^{\infty}_{n=0}\left\{ \frac{(-1)^n}{n!}\frac{\partial^nJ_{ij}^{\alpha}}{\partial r_k^n}\int_{s^{\alpha}}(\hat{r}_k^{\alpha})^nq_j(\bm{y})ds\right\},
\end{eqnarray}
where ${\bm v}^{near}$ is flow due to near-field interaction calculated by Eqs.(\ref{ewald_fla_v}) or (\ref{ewald_v_head}), $N^{far}$ is the number of sperm in the far field, and 
\begin{eqnarray}
\frac{\partial^nJ_{ij}}{\partial r_k^n}{\rvert}_{\bm{y}=\bm{y}^{\alpha}} = \frac{\partial^nJ^{\alpha}_{ij}}{\partial r_k^n}.
\end{eqnarray}
In this study, force-free condition is described as a force-dipole acting on the flagellum and the head.
Therefore, the leading order in the flagellum only is the force term;
\begin{eqnarray}
F_j^{\alpha} = -\int_{s^{\alpha}} q_j(\bm{y})ds.
\end{eqnarray}
Using the multipole expansion method, the flow field is described as follows:
\begin{eqnarray}
v_i = v^{near}_i + \frac{1}{8\pi\mu}\sum^{N^{far}}_{\alpha}\left(J_{ij}^{\alpha}F_j^{\alpha} - J_{ij}^{\alpha}F_j^{\alpha, head}\right).
\end{eqnarray}
A threshold value of $r_c = 3L$ was used to separate the near field from the far field.
\subsection{Numerical procedure}
\ To calculate the swimming of sperm cells, the flagellum is discretized by 30 linear elements.
The elastic force ${\bm q}_e$ is calculated by solving Eq.(\ref{W2}) using a finite-element method coupled with a Gaussian numerical integration scheme.
Then, ${\bm q}_a$ and ${\bm q}_e$ are substituted into Eq.(\ref{2_eq13}), and the hydrodynamic force ${\bm q}_f$ is determined.
The force on the head are computed by Eq.(\ref{F_int}).
The velocities of the tail and head are calculated by Eqs.(\ref{fla_v}) and (\ref{head_v}), respectively, by using a boundary element method [18].
All material points are updated by the second order Runge-Kutta method.
The time step $\Delta t$ is set to $f_0\Delta t = 2.0 \times 10^{-2}$.
We numerically confirmed the time and mesh convergence, and the results did not change considerably for a finer mesh or time step.
To track the orientation of the body frame, we solve a time-change of the orientation vector: 
\begin{eqnarray}
\label{omega}
\frac{d{\bm e}_i}{dt} &=& \omega \times {\bm e}_i,\\
\label{omega2}
\omega &=& \nabla \times {\bm v},
\end{eqnarray}
where ${\bm e}_i$ is the orientation vector of the $i$-th sperm.
The angular velocity $\omega$ is calculated by a finite difference method, and the Runge-Kutta method is also applied to the time-marching of Eq.(\ref{omega}).
{${\bm v}$ in Eq.(\ref{omega2}) is velocity of flow field around the head. To determine the vorticity, we used a finite difference method with $\Delta x/L=0.001$.}

\subsection{Problem Setting}
\begin{figure}[h]
\begin{center}
\resizebox{\textwidth}{!}{
\includegraphics[]{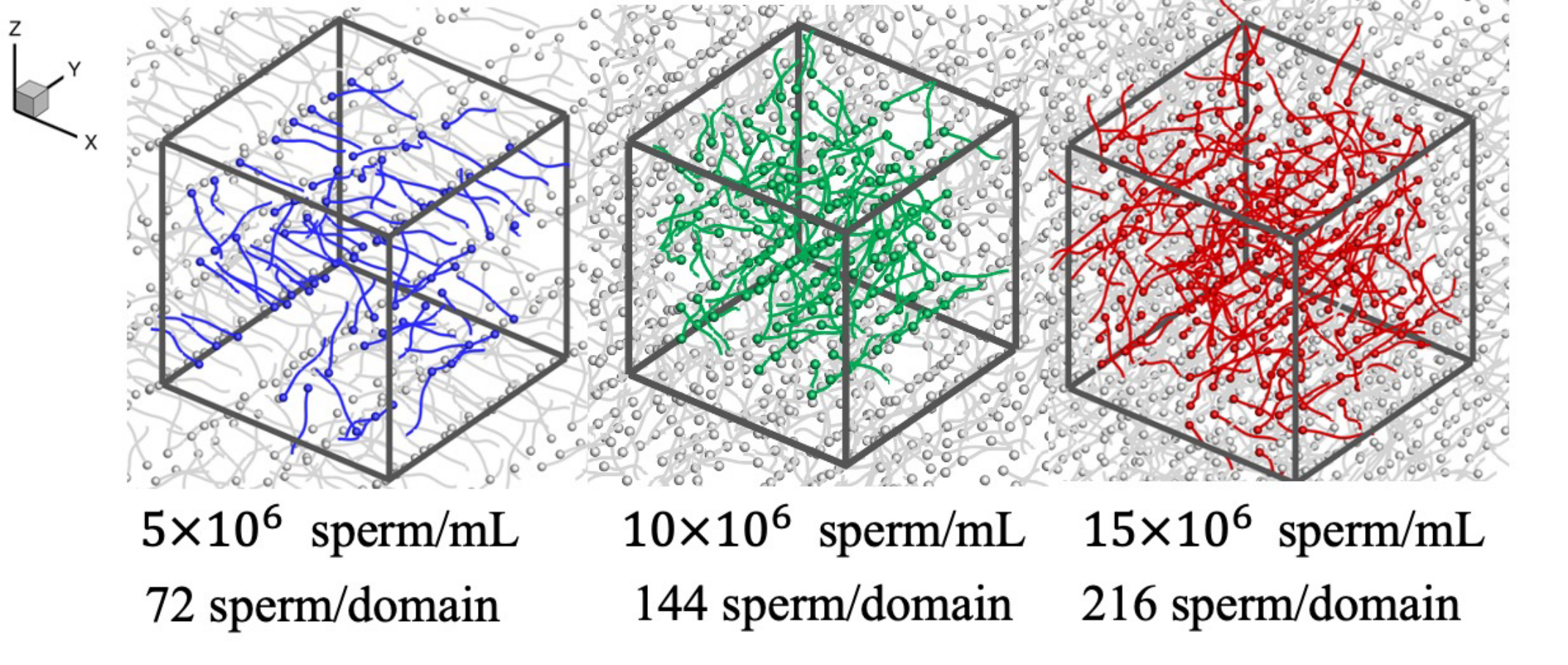}}
\caption{This figure shows each snapshots of low, middle and high concentration with initial condition, $t/T = 0 [-]$ . }
\label{ps}
\end{center}
\end{figure}

As written in Main text, we changed the number of sperm in the main domain to change sperm concentration.
Fig.\ref{ps} shows snapshots of each conditions at $t/T = 0 [-]$.
\\
We also investigated the effect of domain size by changing the domain length from $L_c = 4.0L$ to $L_c = 6.14L$.
The number of sperm is fixed to $N = 260$ for both cases, then the number densities are equivalent to 18 and 5 million sperm/mL, respectively.
Polar ordering was observed to occur under both conditions, but the timescale for reorientation depended on the domain size.
Increasing the domain size also increases the computational cost and makes it difficult to extend the parameter space.
These are outside the scope of this study and will be clarified by future research.
\clearpage
\section{Additional Result}
\subsection{Order parameter}
\begin{figure}[h]
\resizebox{0.51\textwidth}{!}{
\includegraphics[]{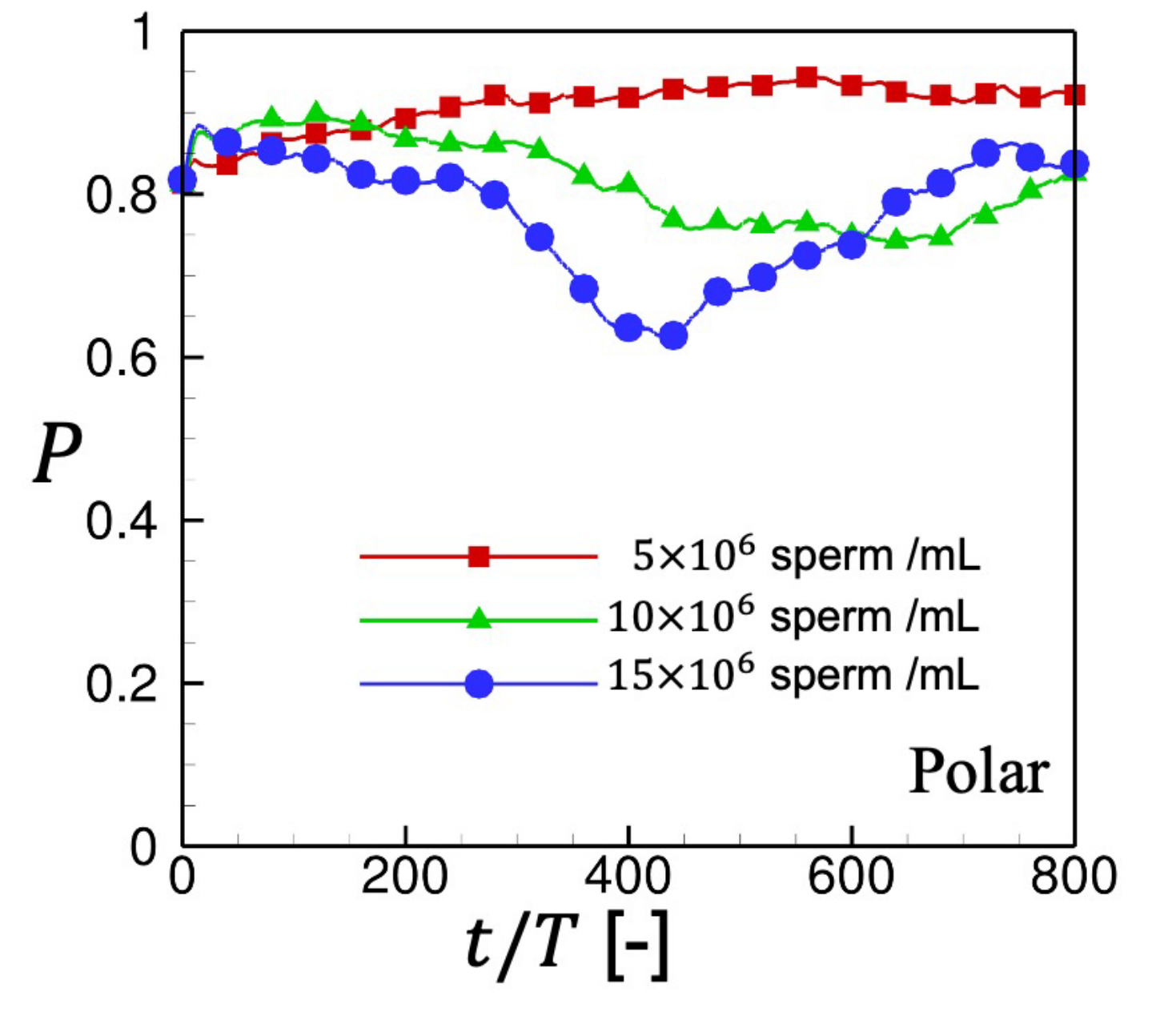}}
\caption{Time change of the orientation in initially polar order}
\label{polar_P}
\end{figure}

\begin{figure}[h]
\resizebox{0.5\textwidth}{!}{
\includegraphics[]{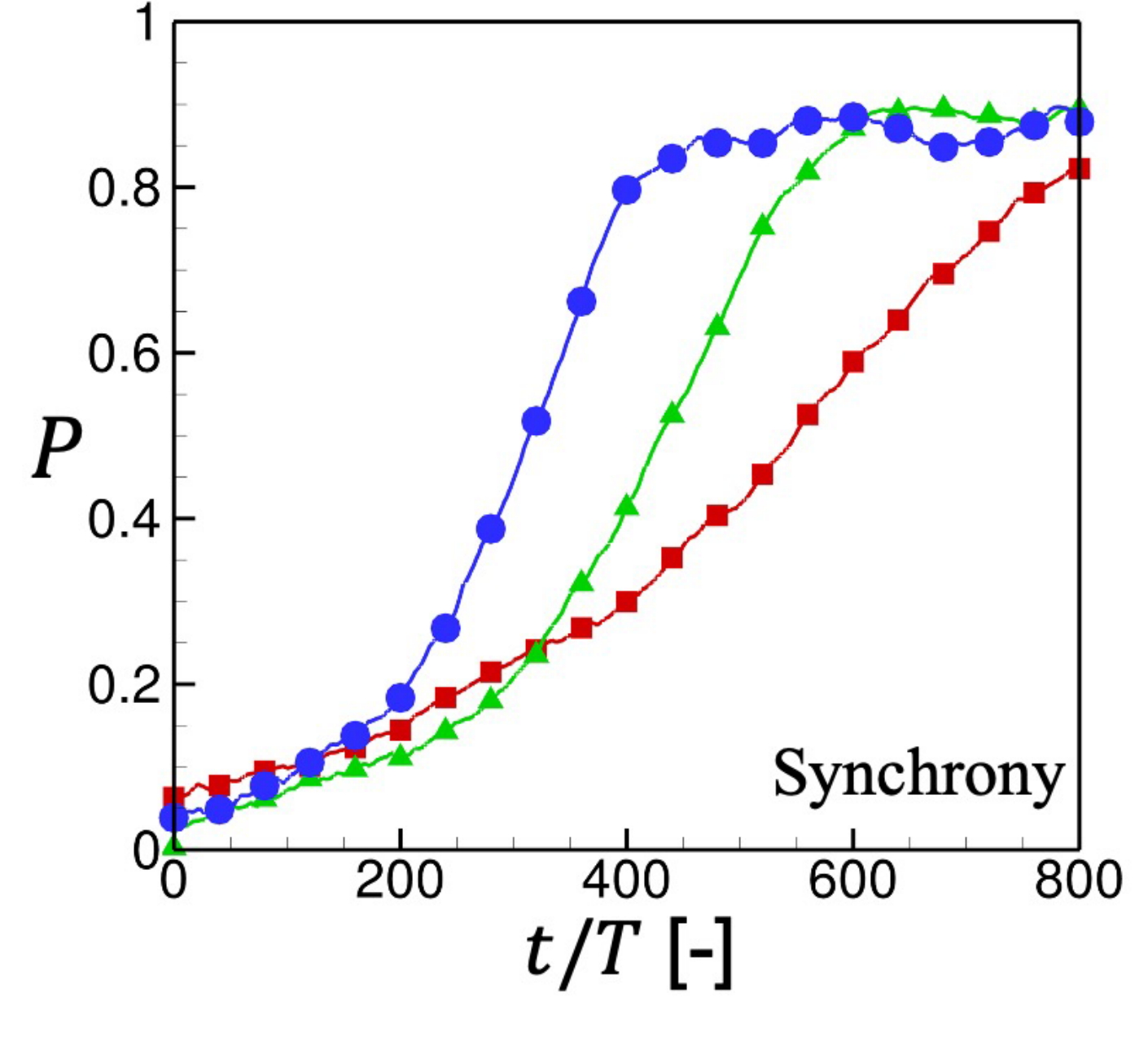}}
\caption{Order parameter $P$ of the random order with self-synchronisation. Three different lines are the results of the same conditions as Fig.\ref{polar_P}.}
\label{geo_P}
\end{figure}

\begin{figure}[h]
\resizebox{0.8\textwidth}{!}{
\includegraphics[]{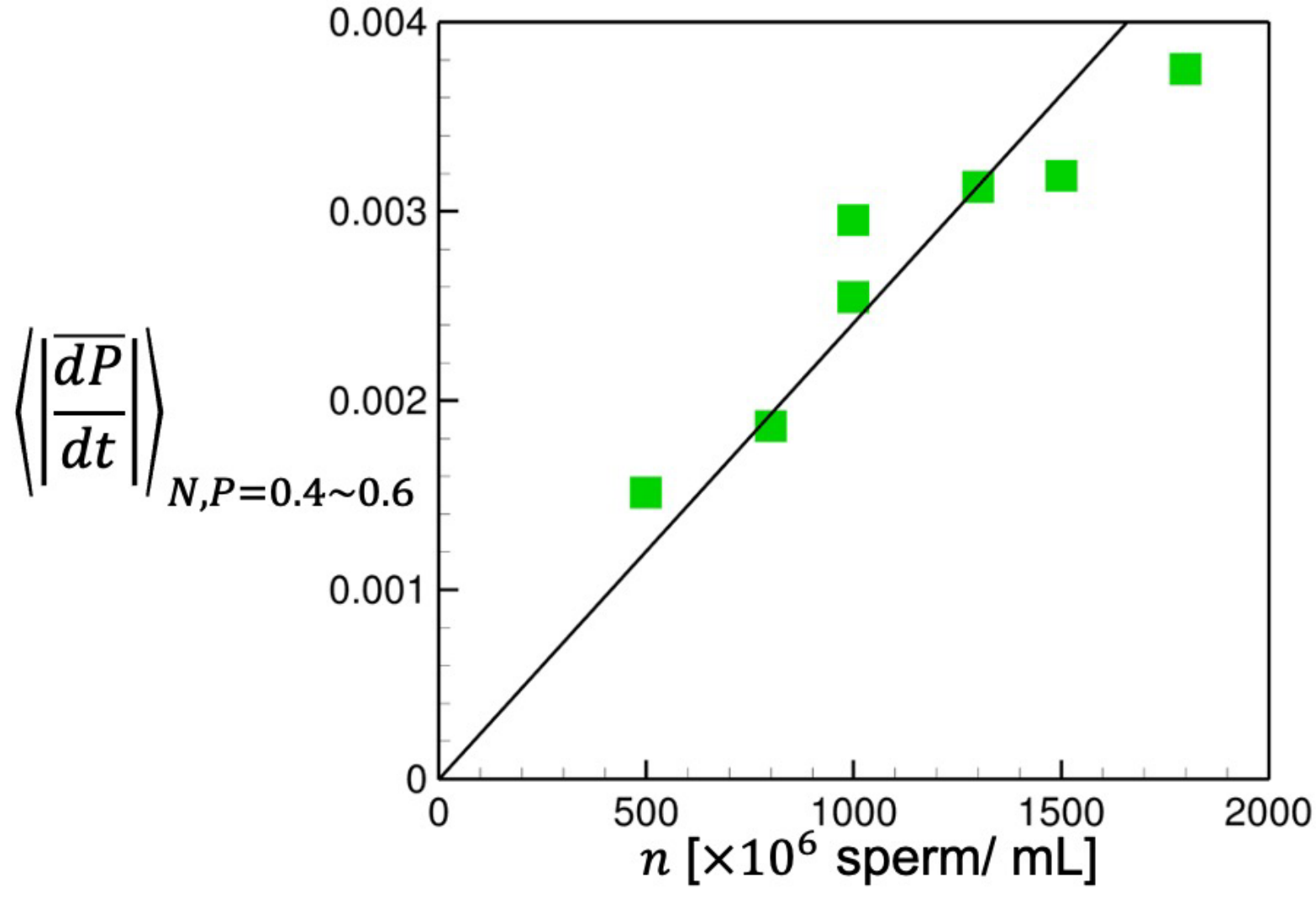}}
\caption{Average orientation changing rate with initially random order}
\label{dP_dt}
\end{figure}

In the main text, we show the time change of the order parameter $P$ in initially random order with various sperm concentration.
In Figs.\ref{polar_P}, and \ref{geo_P}, the effect of initial condition and self-synchronisation on $P$ are shown.
In addition, average of $dP/dt$ at $P=0.4\sim0.6$ with initially random order is shown in Fig.\ref{dP_dt}.


\subsection{Stresslet}
In the main text, we showed stresslet along the average of swimming direction.
The stresslet is defined as
\begin{eqnarray}
\label{stress}
S_{kj} = \frac{1}{2}\int_l \{ \hat{r}_k^{\alpha}q_j (\underline{y})+\hat{r}_j^{\alpha}q_k (\underline{y})-\frac{2}{3}\delta_{jk} \hat{r}_l^{\alpha}q_l(\underline{y})\}.
\end{eqnarray}
To transform the coordinates along the swimming direction, a transformation tensor ${\bm Q}$ is defined, expressed as
\begin{eqnarray}
\label{Q}
Q_{ij} = {\bm e}_i \cdot {\bm g}_j
\end{eqnarray}
where ${\bm e}_i$ is the base vectors in the orthonormal swimmer frame (e.g. ${\bm e}_1$ is the swimming direction), and ${\bm g}_i$ is the base vectors in the Cartesian coordinate. 
Stresslet is then coordinate transformed as
\begin{eqnarray}
\label{transform}
{\bm  S}' =  {\bm Q} \cdot {\bm  S} \cdot {\bm Q}^{\top}.
\end{eqnarray}

\begin{figure}[h]
\resizebox{\textwidth}{!}{
\includegraphics[]{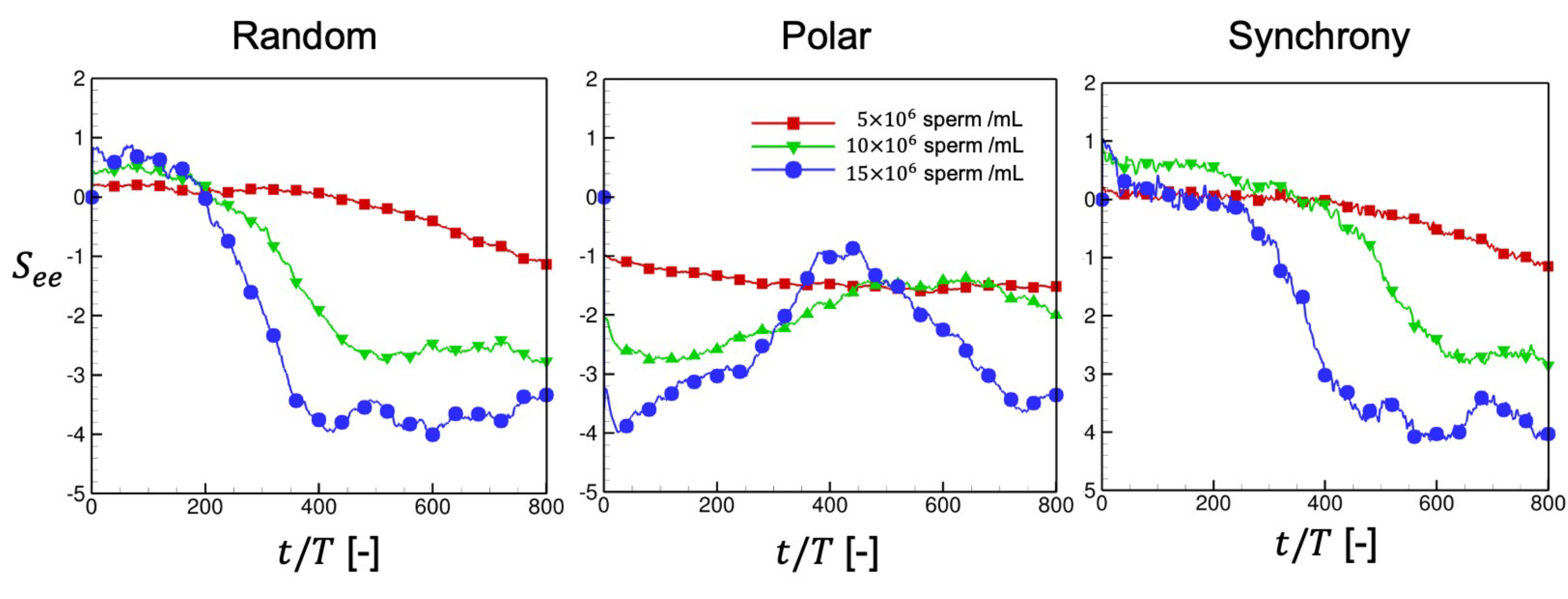}}
\caption{Time change of the stresslet in each initial condition}
\label{See_all}
\end{figure}
\newpage
\subsection{Structure of sperm suspension}
\begin{figure}[h]
\resizebox{0.5\textwidth}{!}{
\includegraphics[]{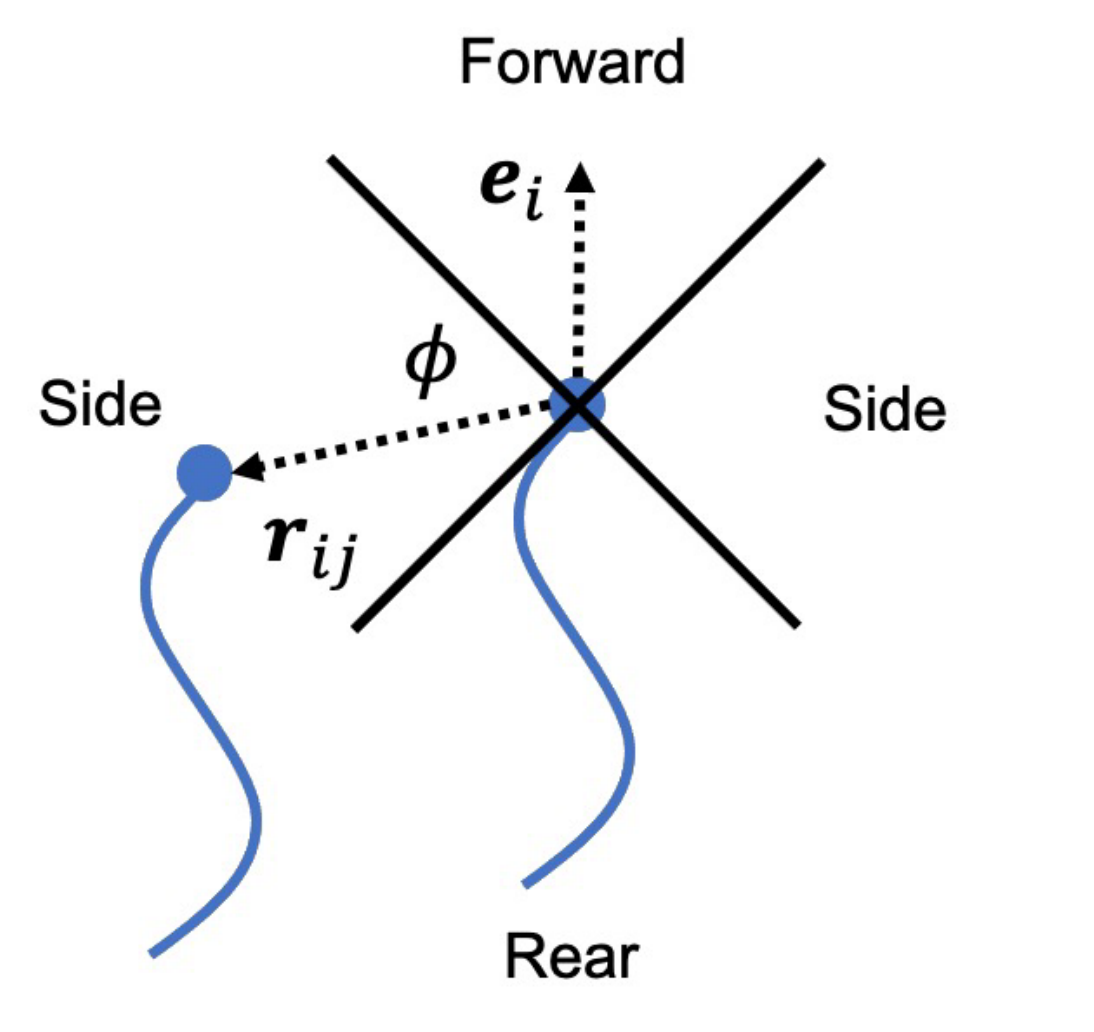}}
\caption{Definition of relative position $i$- and $j$-th sperm: side area or forward-rear}
\label{fore_side}
\end{figure}

\begin{figure}[h]
\resizebox{\textwidth}{!}{
\includegraphics[]{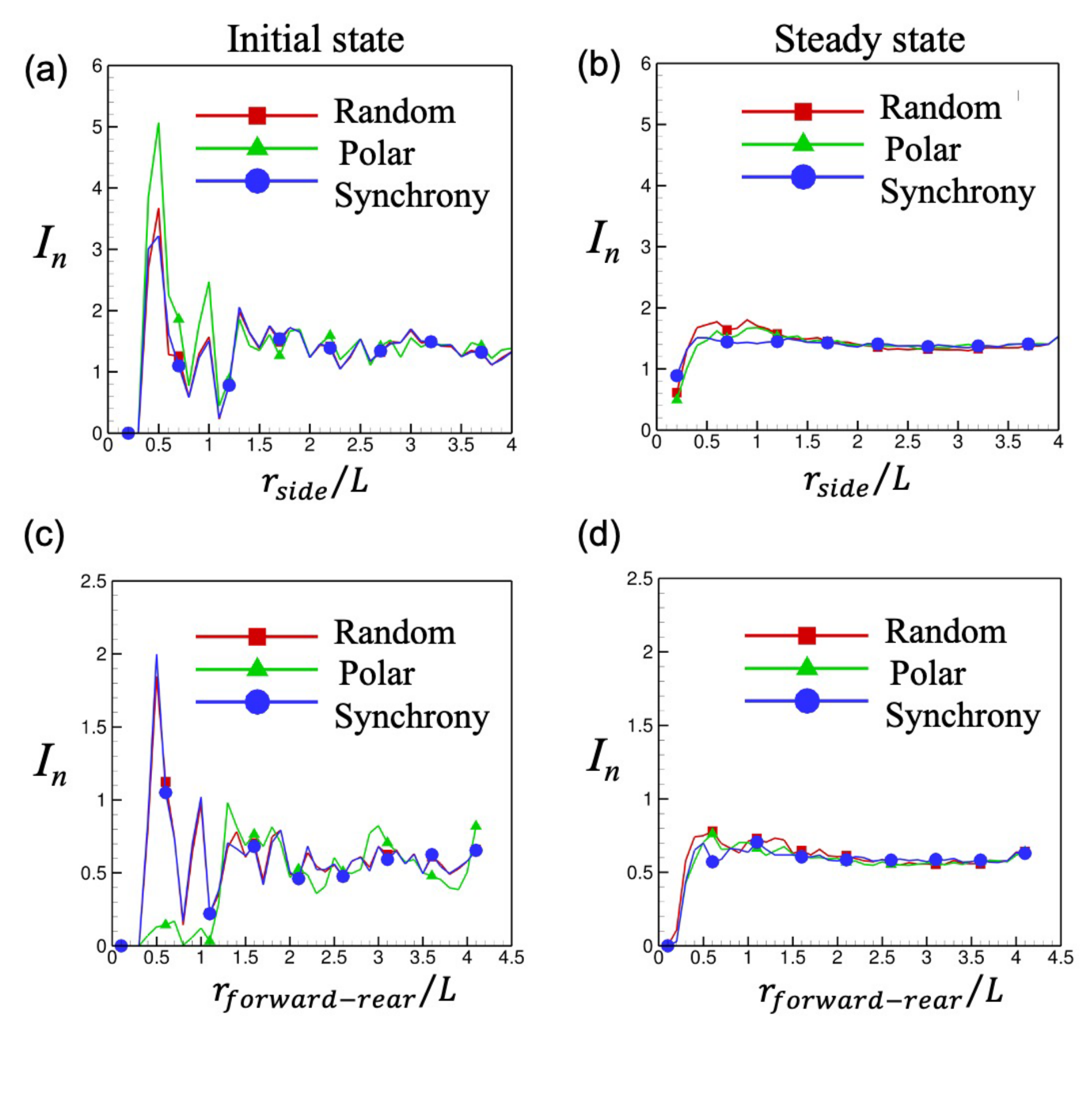}}
\caption{Distance correlated sperm cohesiveness: (a-b) side direction and (c-d) forward-rear direction}
\label{In_fore}
\end{figure}

We calculated cohesiveness $I_n$  to analyze the structure of sperm suspension.
$I_n$ is defined as
\begin{eqnarray}
\label{In}
I_n = \frac{n_{sim}}{n_{ave}},\\
n_{sim} = \frac{n(r_g)}{dV(r_g)},\\
n_{ave} = \frac{N^{all}}{V},
\end{eqnarray}
where $r_g$ is the distance from a sperm and $N(r_g)$ is the number of sperm in the shell volume $dV$, and $n_{ave}$ is the average number density in the domain.
Then, $I_n$ expresses the ratio of the number density in the spherical shell region $r_g$ away from a sperm to the number density of the entire domain.
Also consider whether other sperm are in the forward-rear or lateral direction to their own swimming (cf. Fig.\ref{fore_side}).
Spermatozoa initially arranged in a crystalline structure were found to adopt a homogeneous cell structure by free swimming (cf. Fig.\ref{In_fore}).
\begin{eqnarray}
\label{side_calc}
\phi &=& {\bm e}_i \cdot \frac{{\bm r}_{ij}}{\lvert{\bm r}_{ij}\rvert}\\
{\bm r}_{ij} &=& {\bm x_{head}}_j - {\bm x_{head}}_i\\
\lvert \phi \rvert &\geq& \frac{1}{\sqrt{2}}:\rm{side} \nonumber\\
\lvert \phi \rvert &<& \frac{1}{\sqrt{2}}:\rm{forward-rear}. \nonumber
\end{eqnarray}

\begin{figure}[h]
\resizebox{0.6\textwidth}{!}{
\includegraphics[]{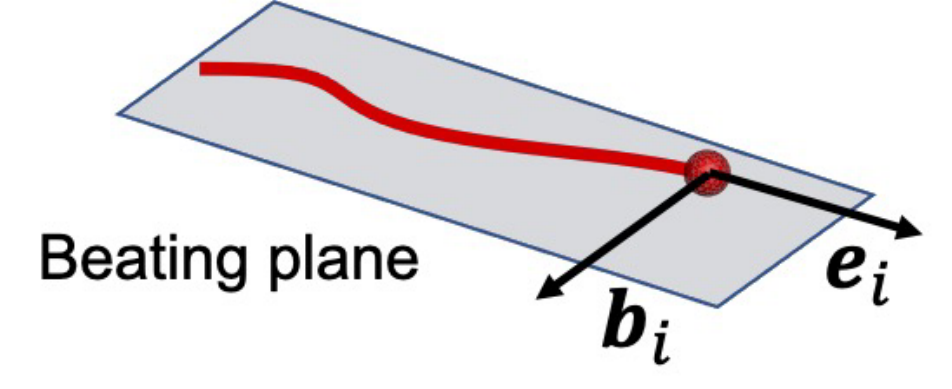}}
\caption{Schematic illustration of main beating plane}
\label{beating}
\end{figure}

\begin{figure}[h]
\resizebox{0.8\textwidth}{!}{
\includegraphics[]{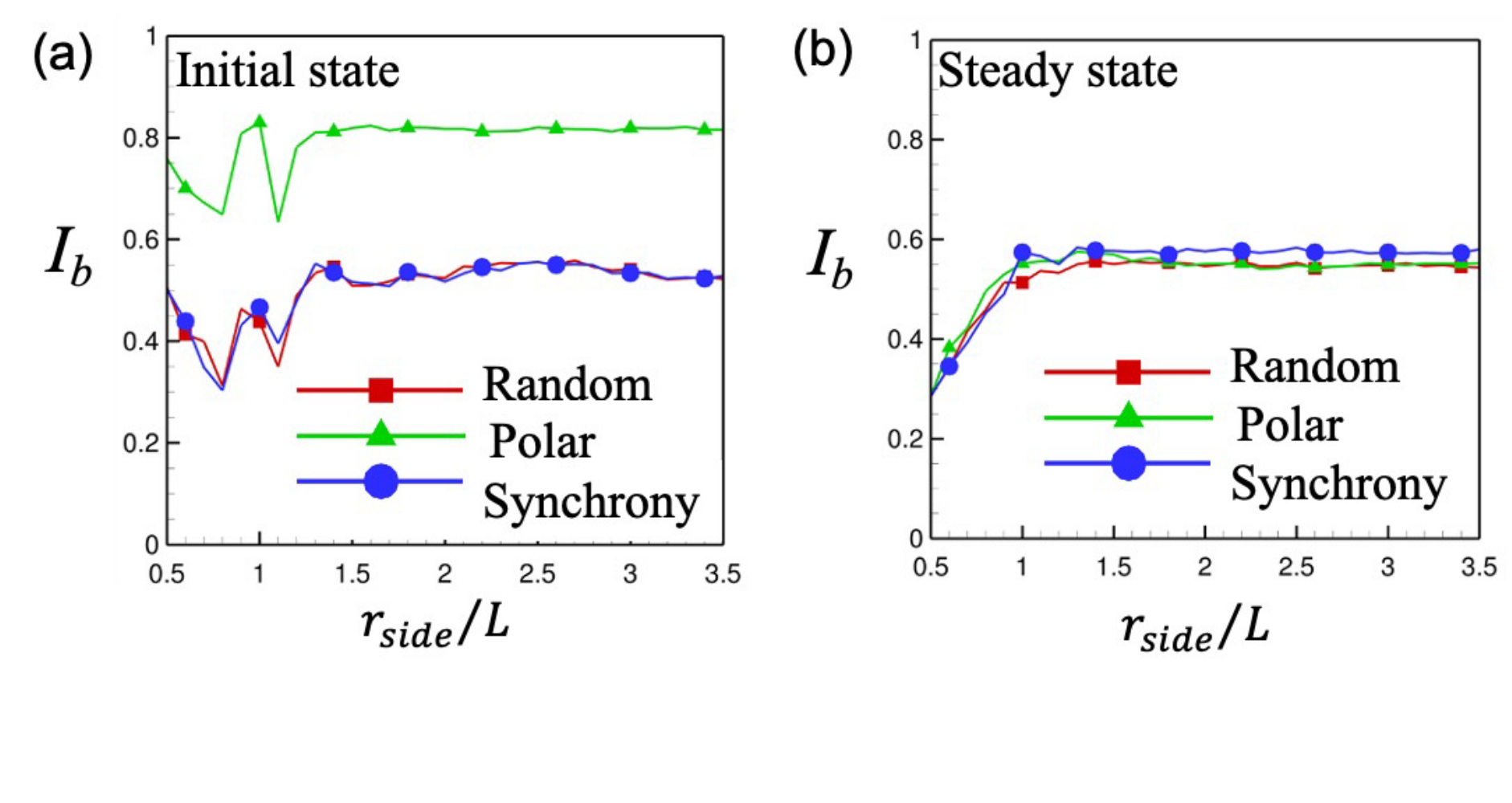}}
\caption{Distance correlations of the main flagellar beating plane}
\label{e2_side}
\end{figure}

The distance correlations of the main flagellar beating plane were also analysed.
The main flagellar beat plane of $i$-th sperm is defined as ${\bm b}_i$ (cf. Fig.\ref{beating}), and the distance correlation is given by
\begin{eqnarray}
I_b (r_{side}, t)= \langle \frac{1}{N_{side}}\sum_j^{N_{side}}\lvert \bm{b}_i\cdot\bm{b}_j\rvert\rangle.
\end{eqnarray}
Results of the correlations in the initial state and steady state are shown in Fig.\ref{e2_side}.
After long time duration, the value became around 0.5 in all cases, whereas it reaches over 0.8 in initially polar order. 
This means that the beating direction of all sperm suspension became almost homogeneous at $r_{side}\ge L$. 

\subsection{Velocity, amplitude, work and swimming efficiency}
\begin{figure}[h]
\resizebox{\textwidth}{!}{
\includegraphics[]{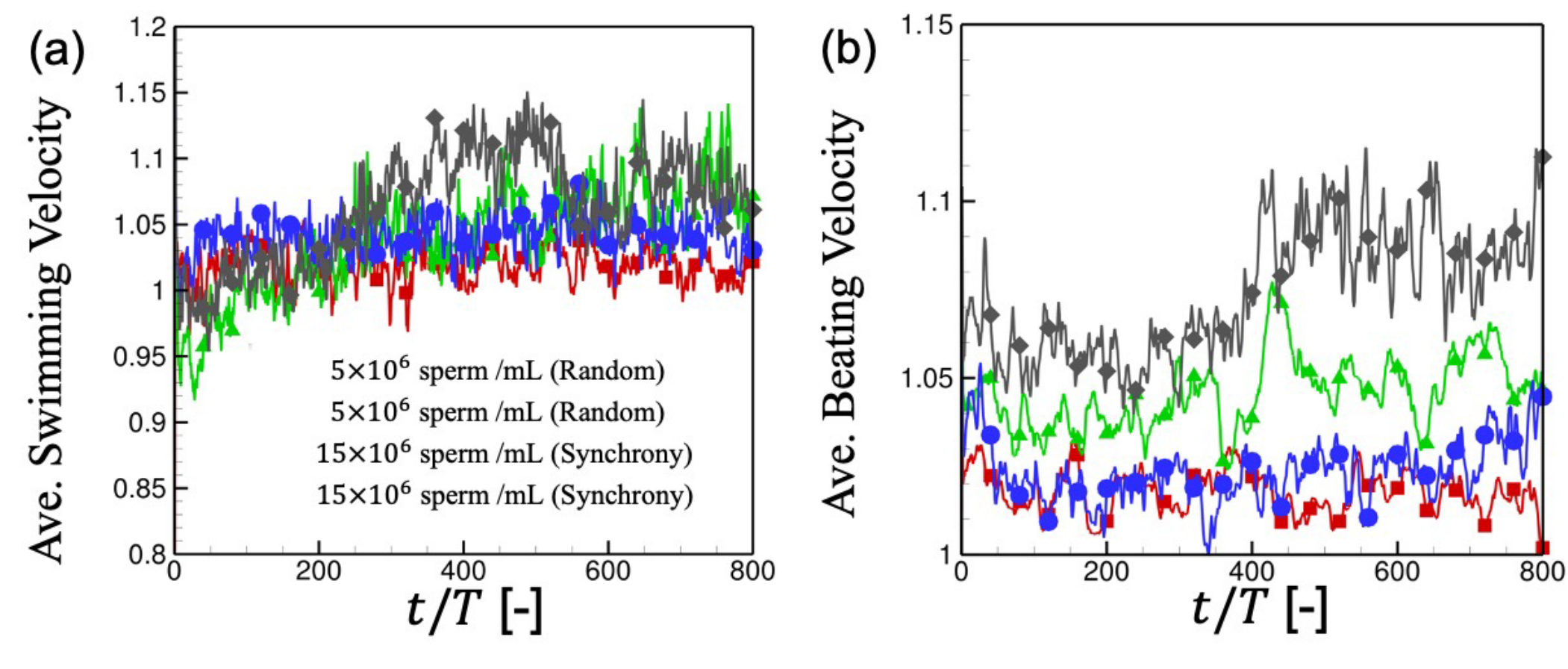}}
\caption{Average swimming velocity and flagellar beating velocity with various conditions. These values are normalised by solitary swimming sperm.}
\label{velocity}
\end{figure}

\begin{figure}[h]
\resizebox{0.6\textwidth}{!}{
\includegraphics[]{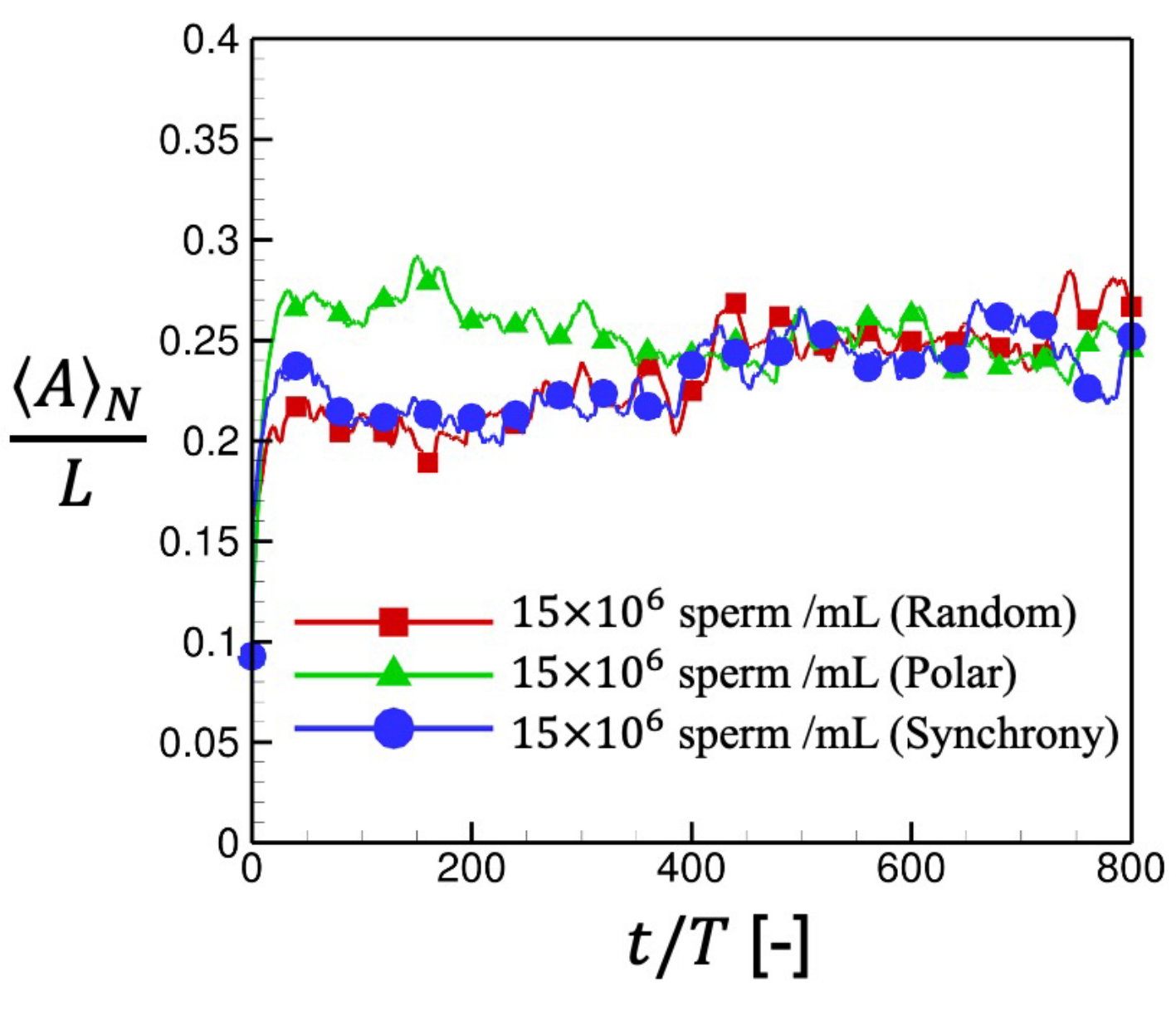}}
\caption{Average amplitude of flagellar beat}
\label{amplitude}
\end{figure}

We also calculated swimming velocity, beat amplitude, work rate, and swimming efficiency.
First, we define swimming velocity as time change of head position: ${\bm U} = d{\bm x}^{head}/dt$, and its average is given by $\left<U\right> = \frac{1}{N}\sum_i^N\lvert d{\bm x}_i^{head}/dt\rvert$.
We also define average flagellar beating velocity: $\langle v^{fla}\rangle = \frac{1}{N}\sum_i^N \frac{1}{L}\int \lvert{\bm v }\rvert ds$.
The results show that, with self-synchronisation, beating velocity can be increased but swimming velocity is less changed (cf. Fig.\ref{velocity}).
The increase in the speed of the beats is not due to an increase in amplitude (cf. Fig.\ref{amplitude}), but due to an increase in frequency, and the reason for the lack of an increase in swimming speed stems from the periodic boundary.

\begin{figure}[h]
\resizebox{0.9\textwidth}{!}{
\includegraphics[]{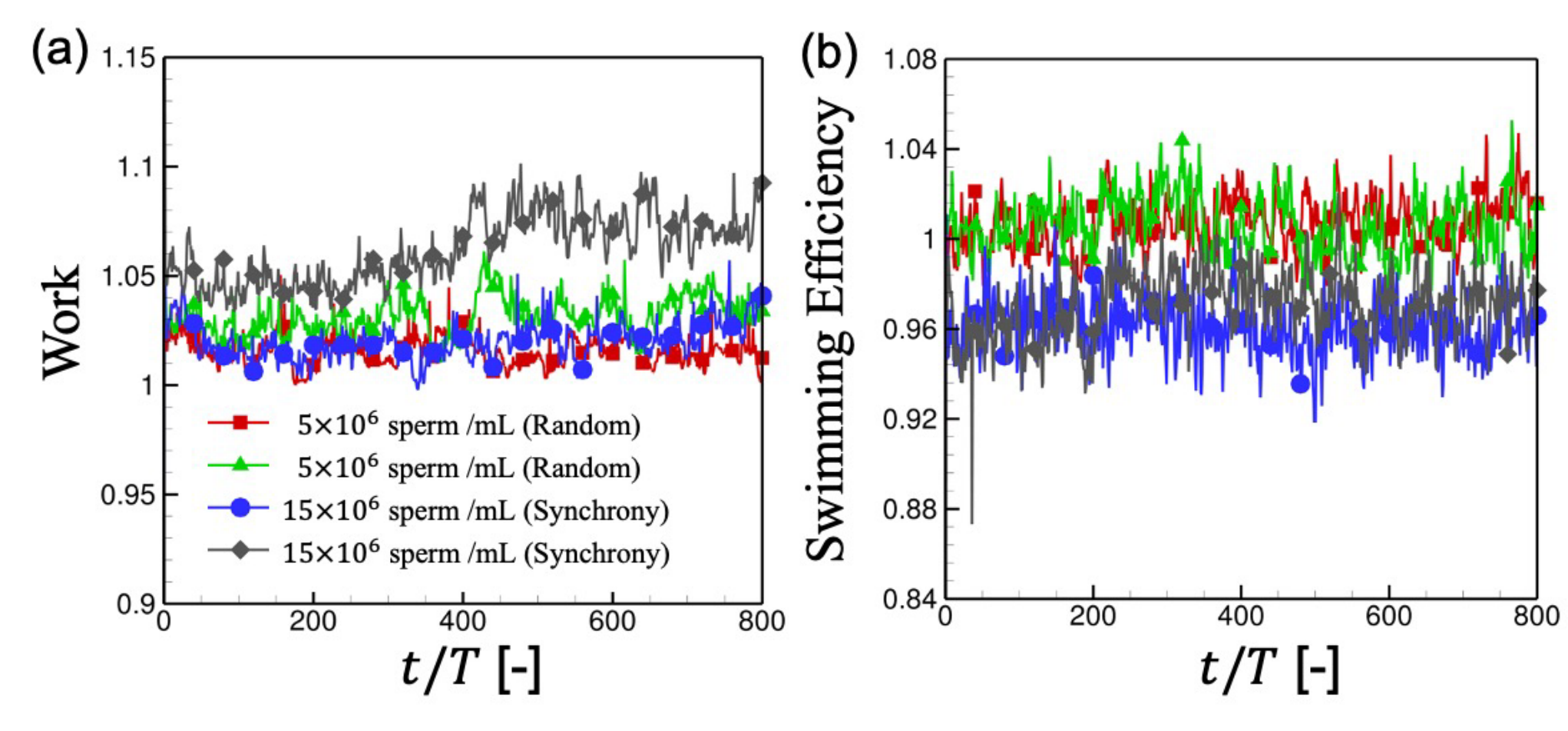}}
\caption{Work rate and swimming efficiency with several concentration conditions}
\label{work}
\end{figure}

Next, we investigate the work rate and swimming efficiency.
We define the work rate due to sperm swimming as follows:
\begin{eqnarray}
\label{work_c}
W &=& W_{fla}+W_{head} = -\int {\bm v}\cdot{\bm q_f}dl+{\bm U} \cdot {\bm F}_{head}.
\end{eqnarray}
Accordingly, the swimming efficiency is defined as follow:
\begin{eqnarray}
\label{eff}
\eta &=& \frac{{\bm U}\cdot {\bm F}_{head}}{W}.
\end{eqnarray}
As shown in Fig.\ref{work}, the beat speed becomes larger due to self-synchronisation, and associated work rate also can be increased.
However, the swimming speed does not change so much, and the efficiency becomes less by introducing self-synchronisation in the infinite suspension.
These trends may change with changes in 3D periodic boundary conditions.

\clearpage
\section{Average phase difference in homogeneous distribution}
In this section, we analytically derive the average phase difference when they are homogeneously distributed.
Assume a state in which the phase of each sperm is uniformly random.
Continuously spreading phase difference is expressed discretely and the phase of each sperm is defined as $\psi_i = i \times \Delta\psi$, where $i$ is the sperm ID, $\Delta\psi = 2\pi/(N-1)$, and $N$ is the number of sperm in the suspension.
The combination of phase difference  magnitude corresponding to each sperm ID is shown in the following table.

\begin{table}[htb]
\centering
  \caption{Combination of phase difference magnitude corresponding to each sperm ID }
  \begin{tabular}{| c| |c|c|c|c|c|c|c|c|c|}  \hline
       & 1 	     	         & 2 		        & 3 		    	& ... & $N$\\ \hline \hline
   1   & 0                    & $\Delta\psi$  & 2$\Delta\psi$ & ...  &$(N-1)\Delta\psi$\\ \hline
    2   & $\Delta\psi$   & 0	 	         & $\Delta\psi$	& ... &$(N-2)\Delta\psi$ \\ \hline
    3   & 2$\Delta\psi$ & $\Delta\psi$  & 0			& ... &$(N-3)\Delta\psi$\\ \hline
    ...  &...			 & ...			& ...			&...	&...\\ \hline
  $N$ &$(N-1)\Delta\psi$&$(N-2)\Delta\psi$&$(N-3)\Delta\psi$&... & 0\\ \hline
  \end{tabular}
\end{table}
From the above combinations, the number of sperm pairs with each phase difference is as follows:
\begin{table}[htb]
\centering
  \caption{Number of pairs with each phase difference}
  \begin{tabular}{|c|c|}  \hline
  Phase difference & Number of pairs\\ \hline
  0 			    & $N$\\
  $\Delta\psi$        &$2(N-1)$\\ 
  2$\Delta\psi$      & $2(N-2)$ \\ 
       ...  	            &...\\ 
 $(N-1)\Delta\psi$ & 2\\ \hline
  \end{tabular}
\end{table}

Total of pairs is $N^2$, and the average expected value $\langle\Delta\psi\rangle$ is given by
\begin{eqnarray}
\langle\Delta\psi\rangle = \frac{N\times0+2\times\sum_{i=1}^{N-1}(N-i)\times i\Delta\psi}{N^2}.
\end{eqnarray}
Substituting $\Delta\psi = 2\pi/(N-1)$, we get
\begin{eqnarray}
\frac{\langle\Delta\psi\rangle}{2\pi}&=&\frac{2}{N^2(N-1)}\sum_{i=1}^{N-1}N\times i-i^2\nonumber\\
&=&\frac{2}{N^2(N-1)}\lbrace\frac{(N-1)N^2}{2}-\frac{(N-1)N(2N-1)}{6}\rbrace\nonumber\\
&=&\frac{(N-1)N(N+1)}{3(N^3-N^2)}\nonumber\\
&=&\frac{N+1}{3N}
\end{eqnarray}
Take N to the limit of infinity, the expected value can be derived as 1/3:
\begin{eqnarray}
\lim_{N \to \infty}{\langle\Delta\psi\rangle}/2\pi = \lim_{N \to \infty}\frac{N+1}{3N} =\frac{1}{3}.
\end{eqnarray}
\section{Sperm density and relaxation time for reorientation}
Finally, we discuss the relationship between sperm density and relaxation time for reorientation.
Sperm has an elongated shape and tends to orient in the direction of elongational flow.
An individual sperm exerts stresslet, which induces the elongational flow in the direction of sperm orientation.
When the orientation of sperm is anisotropic, the elongational flow does not cancel out and make other sperm to align in the elongational direction.
The speed of the elongational flow is roughly proportional to the number of oriented sperm, because the elongational flow in Stokes flow regime is basically induced by a sum of individual stresslets.
Therefore, the alignment rate of sperm becomes proportional to the number of oriented sperm, and the sperm alignment increases exponentially in time.
This tendency was confirmed in Fig. \ref{dP_dt}.
In other words, the order parameter $P(t)$ is expected to increase exponentially in time $t$.
\begin{figure}[h]
\resizebox{0.8\textwidth}{!}{
\includegraphics[]{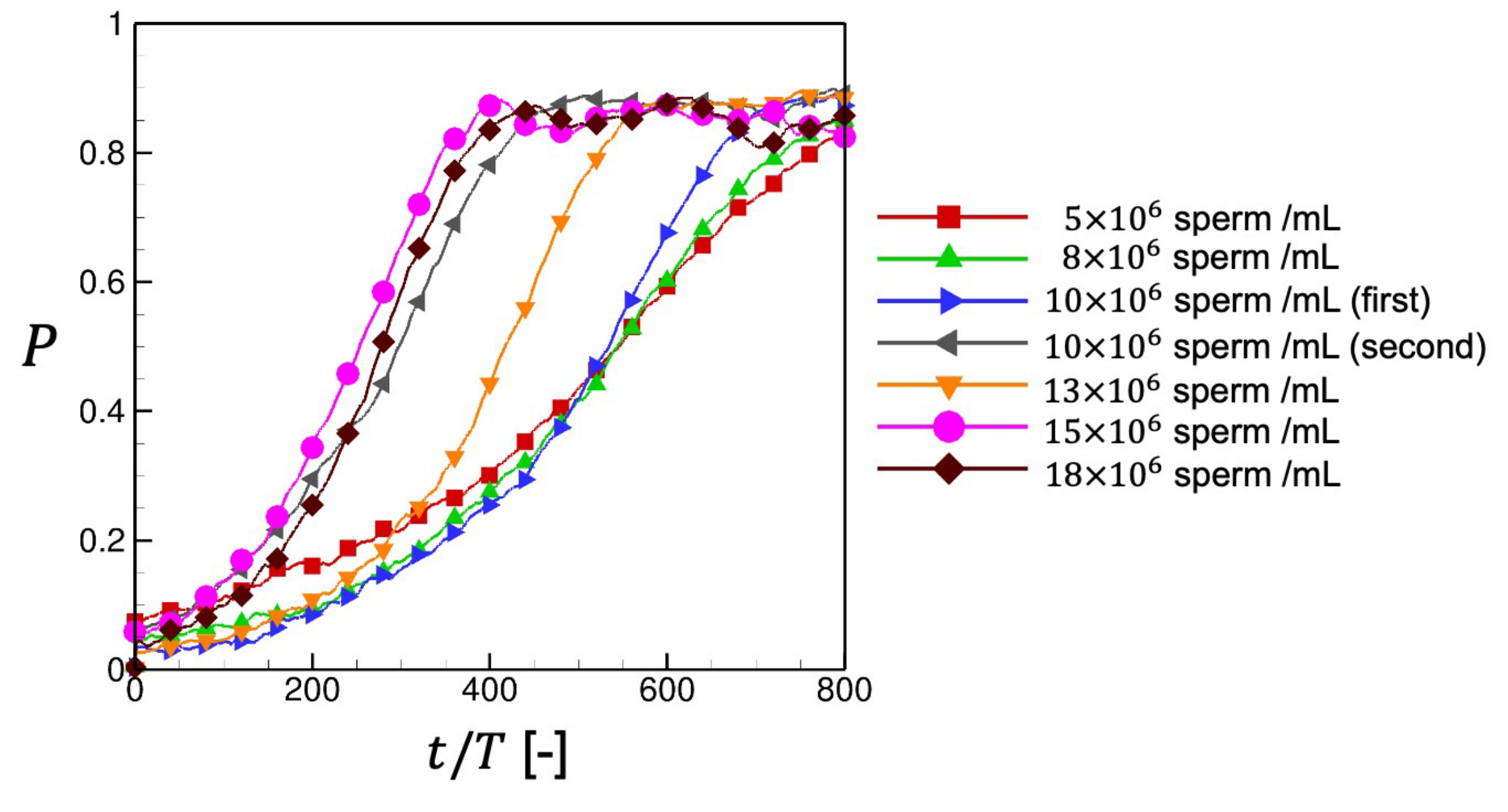}}
\caption{Order parameter $P$ with various sperm density}
\label{P_all}
\end{figure}

\ As we see in Fig.\ref{P_all}, the order parameter $P(t)$ starts with a value greater than 0 and never reaches 1.
The reasons for the incomplete orientation at large $t$ are that the sperm orientation fluctuates during the flagellar beat cycle and is temporarily altered by near-contact interactions between sperm.
At $t=0$, $N$ sperm are placed in a random orientation, and expected value of $P(t=0)$ is $1/\sqrt{N}$ [35].
Given these tendencies, change the definition region for the time scale of $P$ from 0.1 and 0.8, rather than from 0 to 1.

\begin{figure}[h]
\resizebox{0.8\textwidth}{!}{
\includegraphics[]{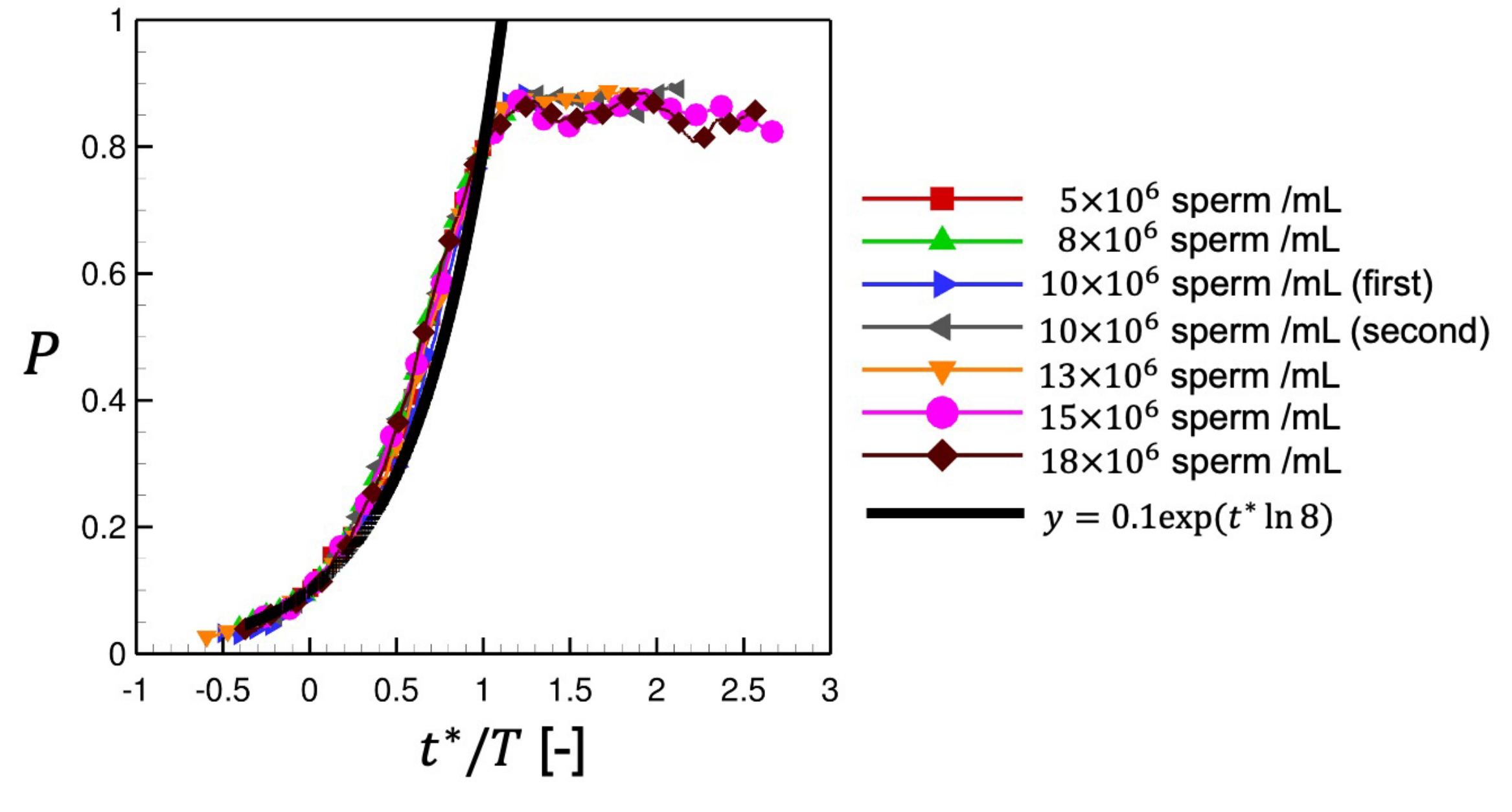}}
\caption{Normalised order parameter}
\label{P_t_new}
\end{figure}

Let $\tau_{0.1}$ be the time when $P(t) = 0.1$, and $\tau_p$ be the time when $P(t) = 0.8$.
We employ the definition as a normalized time $t^* = (t - \tau_{0.1})/(\tau_p - \tau_{0.1})$, the exponential increase of $P(t^*)$ from 0.1 to 0.8 can be written as
\begin{eqnarray}
\label{app_pt}
P(t^*) = 0.1 \exp \left( t^* \ln{8} \right).
\end{eqnarray}
The time change of $P(t^*)$ is replotted using $t^*$, and the results are as shown in Fig.\ref{P_t_new}.
The exponential curve of Eq.(\ref{app_pt}) is also shown in the figure. We see that all curves collapse into a single curve and fit well with Eq.(\ref{app_pt}). These results clearly indicate that the order parameter $P(t)$ increases exponentially in time in all the cases.

At $t = 0$, not $t^* = 0$, we expect $P(t=0) = 1/\sqrt{N}$. By inserting $1/\sqrt{N}$ to the left side of Eq.(\ref{app_pt}), we can derive initial normalized time $t^*_i$ as
\begin{eqnarray}
t^*_i = \log_8 {10} - \frac{1}{2} \log_8 {N} .
\label{app_ti}
\end{eqnarray}
The normalized time duration from the initial state at $t^*_i$ to the aligned state at $t^* = 1$ is given by $1 - t^*_i$. 
Besides, the alignment rate of sperm becomes proportional to the number of oriented sperm. 
Eventually, we can assume that the time duration ${\tau}_p$ for sperm cells to align their swimming direction is proportional to the normalized time duration and inversely proportional to $N$ as ${\tau}_p \propto (1 - t^*_i) / N$. 
This function can be rewritten as
\begin{eqnarray}
\tau_p \propto \frac{1 - \log_8 {10} + \frac{1}{2} \log_8 {N}}{N}
\approx \frac{\log_8 {N}}{2N} \propto \frac{\ln{N}}{N} .
\label{app_taup}
\end{eqnarray}
The coefficient $1 - \log_8 {10}$ is omitted because it is one order of magnitude smaller than $\log_8 {N}$ when $N$ is large as in the present study. We can confirm that the scaling of $\tau_p \propto \ln{N}/N$ shows reasonable agreement with the simulation results in Fig.1(d) in Main text. 

\clearpage
\onecolumngrid